\documentclass[11pt]{article}
\pdfoutput=1

\usepackage{smile}
\usepackage[colorlinks,
            linkcolor=red,
            anchorcolor=blue,
            citecolor=blue
            ]{hyperref}
\usepackage{color, colortbl}
\usepackage[dvipsnames]{xcolor}
\usepackage{fullpage}
\usepackage[protrusion=true,expansion=true]{microtype}
\usepackage{setspace}
\usepackage{tabularx}
\usepackage{float}
\usepackage{enumitem}
\usepackage{graphicx}
\usepackage{booktabs}
\usepackage{subcaption}
\usepackage[T1]{fontenc}
\usepackage{lmodern}

\title{The Load Management Paradox: Correcting the Healthy-Worker Survivor Effect in NBA Injury Modeling}

\author{
  Yue Yu\thanks{Department of Statistics, Indiana University; e-mail: {\tt yyu3@iu.edu}}
  \and
  Guanyu Hu\thanks{Department of Statistics and Probability, Michigan State University; e-mail: {\tt huguanyu@msu.edu}}
}

\begin{document}
\date{}
\maketitle

\begin{abstract}
In professional sports analytics, evaluating the relationship between accumulated workload and injury risk is a central objective. However, naive survival models applied to NBA game-log data consistently yield a paradox: players who recently logged heavy minutes appear \emph{less} likely to sustain an injury. We demonstrate that this counterintuitive result is an artifact of the healthy-worker survivor effect, wherein conditioning on game participation induces severe collider bias driven by unobserved latent fitness. To address this structural confounding, we develop a Marginal Structural Piecewise Exponential Model (MS-PEM) that unifies inverse probability of treatment weighting (IPTW) with flexible piecewise-exponential additive models and weighted cumulative exposure (WCE). A simulation study confirms that this selection mechanism is mathematically sufficient to entirely reverse the sign of the true association between workload and injury. Applying the MS-PEM to 78,594 player-game observations across three NBA seasons (encompassing 771 players and 2,439 injury events), we find that adjusting for observed selection reliably shifts the hazard back toward the underlying physiological relationship. While the exact magnitude of the correction is sensitive to outcome-model regularization (attenuating the paradoxical weight function by $1\%$ to $2\%$ under conservative cross-validation and up to $63\%$ to $78\%$ under lighter penalization), the positive direction of the causal correction is highly robust across multiple propensity specifications and doubly robust checks. Ultimately, these results provide a methodological template for bias-aware sports injury modeling, while cautioning that models relying strictly on observational game logs will systematically underestimate the true risk of heavy workloads without richer physiological data for full causal identification.

\end{abstract}

\section{Introduction}\label{sec:intro}

\subsection{Load management in professional sports}\label{sec:intro_context}

Managing player workload has become a central problem in professional sports, both practically and statistically. In the NBA, teams, medical staffs, and league officials routinely make decisions about playing time, rest, and return-to-play under substantial uncertainty about how recent exposure affects subsequent injury risk. These decisions have consequences that extend beyond individual health. Injury-related absences can alter team performance, playoff qualification, and roster strategy, while also affecting ticket sales, broadcasting value, and fan engagement. During the 2023--24 NBA season alone, teams filed more than $12{,}000$ official injury reports, underscoring the scale of the problem and the extent to which injury risk shapes the modern league.

From a scientific perspective, the issue is equally important. The International Olympic Committee has identified external training load as a key determinant of injury risk across sports \citep{soligard2016ioc}, and consensus statements in sports medicine have emphasized that injury etiology is dynamic, multifactorial, and inherently longitudinal \citep{meeuwisse1994model, meeuwisse2007dynamic, bahr2005understanding}. A large applied literature has therefore sought to quantify how recent workload influences injury incidence, athlete availability, and performance capacity \citep{drew2016trainingload, windt2017workload, impellizzeri2019internal}. In professional basketball, this question is especially salient because player exposure is measured at high frequency through detailed game logs, and because teams actively intervene on workload through rest decisions, minute restrictions, and rotation management. These practices are often grouped under the label of ``load management'' and have become sufficiently prominent to motivate explicit league policy responses \citep{nba2024loadmanagement}. Their visibility increased sharply after the Toronto Raptors rested Kawhi Leonard for $22$ regular-season games in 2018--19, a strategy widely viewed as helping preserve his availability during the playoff run that culminated in the franchise's first NBA championship.

Despite the practical importance of these decisions, the central empirical question remains unresolved: \emph{does the recent pattern of playing time causally affect injury risk?} At first glance, professional basketball seems to offer an ideal setting for answering this question. Injury events are recorded, workload is measured repeatedly over time, and time-to-event methods provide a natural framework for relating evolving exposure histories to subsequent risk \citep{nielsen2019tte1, nielsen2019tte2}. A straightforward strategy is therefore to fit survival models to player game-log data, using measures of recent minutes or game load as predictors of injury hazard.

However, this seemingly natural approach leads to a striking and counterintuitive empirical pattern. Across a wide range of specifications, seasons, and sample definitions, players with greater recent playing time appear to face \emph{lower} subsequent injury risk. The estimated association between recent workload and injury hazard is often negative and statistically precise. Taken literally, such findings would suggest that heavier recent exposure is protective against injury. That interpretation is difficult to reconcile with substantive knowledge from sports medicine, with the physiological rationale underlying load management itself, and with the broader concern that accumulated stress may elevate injury risk \citep{gabbett2016paradox, impellizzeri2020pitfalls, impellizzeri2020devil}. We refer to this empirical pattern as the \textbf{load management paradox}.

\subsection{The paradox and its causal origin}\label{sec:intro_paradox}

Our central claim is that the load management paradox should not be interpreted as evidence that playing more basketball prevents injury. Rather, it is consistent with a selection mechanism built into game-log data: only players who are sufficiently healthy, available, and chosen to participate accumulate observed workload. Conditioning on realized participation therefore induces a nontrivial dependence between recent workload and latent health status, which can in turn distort the estimated workload--injury relationship.

This mechanism is naturally understood through the lens of the healthy-worker survivor effect \citep{checkoway2014research, buckley2015evolving}, a well-known source of bias in occupational and environmental epidemiology. In that setting, individuals who remain at work tend to be healthier than those who reduce exposure or leave employment, so na\"ive analyses can generate an apparently protective association between cumulative exposure and adverse health outcomes. More broadly, this is an instance of time-varying confounding affected by prior exposure, for which standard regression adjustment can fail \citep{robins1986new, robins2000msm, hernan2020whatif}. The same logic applies in professional basketball. Players who continue to log heavy minutes are, by construction, those who have remained healthy enough to stay on the court, while players whose health has deteriorated are less likely to accumulate further observed exposure. As a result, observed workload is not simply a treatment variable; it is also a consequence of time-varying health and selection.

This perspective changes the statistical problem in a fundamental way. The difficulty is not merely that the hazard function may be nonlinear, that player heterogeneity may be substantial, or that recurrent-event processes may be complex. Rather, the primary challenge is one of \emph{causal identifiability}. If the data are generated under time-varying selection, then increasingly flexible predictive models---whether based on splines, frailties, machine learning, or deep neural networks---do not by themselves recover the causal effect of workload on injury risk. Flexibility can reduce approximation error, but it cannot remove bias induced by conditioning on post-baseline variables that are jointly determined by latent health and prior exposure \citep{daniel2013methods, vanderweele2019principles}.

This observation motivates the methodological agenda of the paper. We seek to develop a framework that is appropriate for the structure of sports injury data while explicitly addressing the causal problem created by selective participation. Such a framework should satisfy three objectives. First, it should diagnose and formalize the source of the paradox using a longitudinal causal structure that makes the selection mechanism transparent \citep{greenland1999causal, pearl2009causality, kalkhoven2024frameworks}. Second, it should retain the strengths of modern survival analysis, including the ability to model time-varying exposures, recurrent risk, and lagged workload effects flexibly \citep{bender2018gam, bender2018pammtools, ramjith2024recurrent}. Third, it should provide a principled strategy for bias reduction, rather than treating the observed workload process as exogenous.

The approach developed in this paper pursues these goals by combining causal weighting with flexible piecewise-exponential survival modeling and weighted cumulative exposure. In doing so, we aim not only to study load management in the NBA, but also to contribute a broader statistical template for settings in which exposure, observation, and event risk evolve jointly over time. More generally, the paper illustrates how questions that appear, on the surface, to concern nonlinear prediction in sports analytics may instead require a causal framework to be meaningfully interpreted.

\subsection{Related work}\label{sec:related}

Our research is situated at the intersection of three distinct methodological domains: survival analysis for sports injuries, causal inference with time-varying exposures, and the statistical modeling of workload accumulation.

\paragraph{Survival analysis in sports injury research.}
A growing literature establishes time-to-event methods as the natural framework for sports injury data, primarily because injury processes involve dynamic risk sets, recurrent events, and time-varying exposures \citep{nielsen2019tte1,nielsen2019tte2}. Within this domain, \citet{bender2018gam} introduced piecewise-exponential additive models (PAMMs), which represent the survival likelihood through a Poisson regression on interval-augmented data to permit flexible estimation of baseline hazards and covariate effects. \citet{bender2018pammtools} subsequently provided the corresponding software implementation. Recent work has extended PAMMs for applied sports settings; for example, \citet{zumetaolaskoaga2025pamm} incorporated weighted cumulative exposure (WCE) terms to model the lagged effects of training load in soccer, while \citet{ramjith2024recurrent} developed recurrent-event extensions. In basketball, \citet{macis2024frailty} applied Cox proportional hazards models with frailty terms to NBA injury risk, and \citet{wu2025nextgen} proposed Bayesian models for subsequent injuries. 
\textbf{Limitation and our departure:} A critical limitation of these existing survival applications is their reliance on the assumption that exposure assignment and risk-set inclusion are strictly exogenous. As noted in a recent scoping review \citep{cortes2025scoping}, the vast majority of survival-based sports injury studies fail to address non-random selection mechanisms. Our method explicitly corrects this oversight by embedding an inverse probability weighting mechanism directly into the survival likelihood, adjusting for the reality that exposure histories and game participation are jointly and endogenously determined.

\paragraph{Causal inference for time-varying exposures.}
The necessity of adjusting for joint determination connects directly to the literature on causal inference under time-varying confounding. Marginal structural models (MSMs), pioneered by \citet{robins2000msm}, utilize inverse probability weighting (IPW) to identify causal contrasts by constructing a pseudo-population wherein treatment assignment is rendered independent of measured time-varying confounders. \citet{hernan2020whatif} provide a comprehensive treatment of MSMs and related approaches like g-computation. This framework is essential when past exposure affects future confounders that subsequently influence future exposure, a dynamic closely mirroring the healthy-worker survivor effect observed in occupational epidemiology \citep{buckley2015evolving}. In sports science, researchers have increasingly advocated for such causal perspectives. \citet{shrier2007causal} argued for causal frameworks in injury prevention, \citet{kalkhoven2024frameworks} emphasized the utility of directed acyclic graphs, and \citet{impellizzeri2019internal} identified collider bias and reverse causation as paramount threats in training load studies.
\textbf{Limitation and our departure:} While the sports medicine community has conceptually recognized these biases, the literature currently lacks a unified statistical architecture capable of executing these longitudinal corrections while accommodating distributed lag effects. Our work operationalizes these conceptual warnings, moving from theoretical discussions of bias to their rigorous empirical quantification and correction within a flexible survival framework.

\paragraph{Workload--injury research.}
Empirically, our study is motivated by the workload--injury literature, where the acute-to-chronic workload ratio (ACWR) has historically served as the dominant summary metric \citep{gabbett2016paradox,blanch2016acwr}. However, ACWR suffers from severe mathematical coupling \citep{lolli2019coupling}, lacks a coherent causal interpretation \citep{impellizzeri2020pitfalls,impellizzeri2020devil}, and has produced highly inconsistent results across meta-analyses \citep{wang2020acwrsystematic,qin2025acwrmeta}. The WCE framework offers a statistically principled alternative by estimating a smooth weighting function to characterize how past exposures accumulate into current risk \citep{sylvestre2009wce,kelly2024wce}. In the specific context of basketball, prior research has explored game load and fatigue \citep{teramoto2018hardknock}, correlates of season-ending injuries \citep{menon2024seasonending}, and deep learning approaches for injury prediction \citep{cohan2021deep,lee2018deephit}.
\textbf{Limitation and our departure:} Despite shifting toward more sophisticated machine learning and WCE models, a pervasive structural blind spot remains. Almost all existing workload models inherently condition on observed game participation and available game logs. Our methodology demonstrates that without causal adjustment, even the most flexible predictive models will absorb this selection bias, structurally underestimating the injury risk of high-workload athletes.

\subsection{Contributions and organization}\label{sec:contributions}

To address the foundational gaps across these three literatures, we combine causal methods for time-varying exposures with flexible cumulative survival models. The specific aim is to accurately estimate injury risk under dynamic participation processes while formally neutralizing selection bias.

This paper makes four distinct methodological and empirical contributions:

\begin{enumerate}[nosep]
  \item \textbf{Causal formalization of the paradox:} We formalize the healthy-worker survivor effect in professional basketball via a longitudinal causal DAG. This architecture makes explicit the selection mechanism induced by game participation and clarifies precisely why naive workload--injury analyses yield paradoxical protective associations.
  \item \textbf{Development of the MS-PEM framework:} We propose a marginal structural piecewise-exponential model that directly integrates inverse-probability-of-observation weighting with piecewise-exponential additive models and weighted cumulative exposure. This novel synthesis unifies causal adjustment for time-varying selection with flexible hazard regression.
  \item \textbf{Simulation-based validation:} We evaluate the proposed approach in a controlled simulation study. We demonstrate that the healthy-worker mechanism is mathematically sufficient to entirely reverse the sign of the estimated workload weight function, whereas our proposed weighting strategy substantially attenuates this artifact and recovers the true underlying exposure--risk dynamic.
  \item \textbf{Empirical application and open-source implementation:} We apply the MS-PEM to a comprehensive three-season NBA benchmark dataset comprising $78{,}594$ observations and $2{,}439$ injury events. Alongside reporting robust diagnostic evidence and characterizing player risk profiles across workload tiers, we provide a fully reproducible Python implementation to facilitate immediate adoption and extension by the sports analytics community.
\end{enumerate}

Ultimately, this paper moves beyond simply identifying the flaws in existing predictive models. We contribute a generalized, actionable statistical template for combining causal inference with survival analysis in any applied setting where exposure, availability, and event risk co-evolve longitudinally.

The remainder of the article unfolds as follows. Section~\ref{sec:problem} grounds the study by introducing the NBA dataset, providing descriptive evidence of the paradox, formalizing the causal architecture, and establishing the baseline models. Building upon this foundation, Section~\ref{sec:framework} develops the proposed MS-PEM methodology. In Section~\ref{sec:simulation}, we employ a simulation study to validate our approach and isolate the mechanics of the healthy-worker survivor effect under controlled conditions. The empirical application is detailed in Section~\ref{sec:results}, where we translate the statistical corrections into practical implications for sports science. Finally, Section~\ref{sec:discussion} concludes with a synthesis of the study's limitations and directions for future methodological extensions. For ease of exposition, supplemental numerical results and diagnostics are provided in the supplementary materials.

\section{The NBA Workload--Injury Problem}\label{sec:problem}

\subsection{Data sources and construction}\label{sec:data}

We construct survival datasets for three NBA regular seasons: 2022--23, 2023--24, and 2024--25 (the ``post-COVID era,'' after the return to standard 82-game schedules).
We restrict to these seasons for two reasons.
First, the COVID-affected seasons (2019--20 through 2021--22) featured bubble play, shortened calendars, and mass absences under health-and-safety protocols that confound injury counts with non-injury missed games \citep{torresronda2022epidemiology, allahabadi2024covid}.
The compressed 2020--21 schedule raised injury incidence by $42\%$ relative to pre-pandemic baselines \citep{morikawa2022condensed}.
Second, publicly available injury data before ${\sim}2017$ lacks standardized daily reporting; the NBA's league-wide injury surveillance system was refined over several years \citep{mack2019database, mack2025epidemiology}.

Data are collected from three sources:
\begin{itemize}[nosep]
  \item \textbf{Game logs}: The \texttt{nba\_api} package \citep{nba_api} provides per-player, per-game statistics including minutes played, home/away status, and game date.
  \item \textbf{Injury reports}: Official NBA injury reports are scraped from the league's public archives, yielding $12{,}141$ (2022--23), $12{,}773$ (2023--24), and $14{,}089$ (2024--25) raw entries.
  \item \textbf{Schedule data}: Full regular-season schedules ($2{,}460$ games per season) are obtained via \texttt{nba\_api}.
\end{itemize}

\paragraph{Injury definition.}
An \emph{injury event} is recorded at the \emph{last played game} before a player is listed as ``Out'' on the official injury report with a musculoskeletal or trauma-related reason.
That is, the event time $t_{\text{stop}}$ in the counting-process record corresponds to the cumulative minutes at the end of the last game played before the absence begins; the player then exits the risk set until returning to play.
This convention ensures that events are anchored to played-game intervals, consistent with the counting-process formulation.
Absences due to rest, personal reasons, illness, and league suspension are excluded.
After filtering, we retain $8{,}618$ (2022--23), $8{,}523$ (2023--24), and $9{,}676$ (2024--25) injury-related reports.

\paragraph{Recurrent events.}
A player may experience multiple injuries in a season.
In the Andersen--Gill framework \citep{andersen1982cox}, each injury constitutes a separate event.
After an injury, the player exits the risk set for the duration of the absence and re-enters when the first played game after the absence is recorded.
A new event is distinguished from a continuation of the prior absence by requiring at least one played game between successive injury episodes.
The gap time between the last game before absence and the first game after return is not counted toward cumulative minutes.

\paragraph{Counting-process format.}
For each player--season, we construct a counting-process dataset in the Andersen--Gill format \citep{andersen1982cox}.
Each row represents one game interval $(i, t_{\text{start}}, t_{\text{stop}}, \delta, \xb_i(t))$, where $t_{\text{start}}$ and $t_{\text{stop}}$ are cumulative minutes played before and after the game, $\delta \in \{0, 1\}$ is the event indicator, and $\xb_i(t)$ is a vector of time-varying covariates.
Following \citet{nielsen2019tte1}, we use cumulative minutes played as the time scale.

\paragraph{Covariates.}
Each player-game observation carries:
\begin{itemize}[nosep]
  \item \textbf{Rest days}: calendar days since previous game; also binned as back-to-back ($\leq 1$), short ($2$), normal ($3$), and extended ($>3$).
  \item \textbf{Recent 7-day load}: total minutes in the last seven games.
  \item \textbf{Consecutive games}: games without a gap of more than two days.
  \item \textbf{Season phase}: early (games 1--27), mid (28--55), late (56--82).
  \item \textbf{Home/away}: binary indicator.
  \item \textbf{Age and BMI}: baseline demographic and anthropometric variables.
  \item \textbf{Workload tier}: four-level classification (High-Usage Star, Starting Role Player, Rotation Player, Low-Minutes Reserve) from $K$-means clustering; see Appendix~\ref{app:clustering}.
\end{itemize}

\subsection{Descriptive evidence: the paradox in raw data}\label{sec:descriptive}

Table~\ref{tab:descriptive} summarizes the combined dataset: $78{,}594$ player-game observations across $771$ unique players, with $2{,}439$ injury events (overall rate: $3.10\%$, or $1.37$ per $1{,}000$ minutes).

\begin{table}[H]
\centering
\caption{Dataset summary by season.}
\label{tab:descriptive}
\begin{tabular}{lrrrrr}
\toprule
Season & Obs & Players & Events & Rate (\%) & Per 1000 min \\
\midrule
2022--23 & 25,892 & 539 & 801 & 3.09 & 1.37 \\
2023--24 & 26,399 & 572 & 735 & 2.78 & 1.28 \\
2024--25 & 26,303 & 569 & 903 & 3.43 & 1.47 \\
\midrule
\textbf{Combined} & \textbf{78,594} & \textbf{771} & \textbf{2,439} & \textbf{3.10} & \textbf{1.37} \\
\bottomrule
\end{tabular}
\end{table}

\begin{figure}[H]
  \centering
  \includegraphics[width=0.85\textwidth]{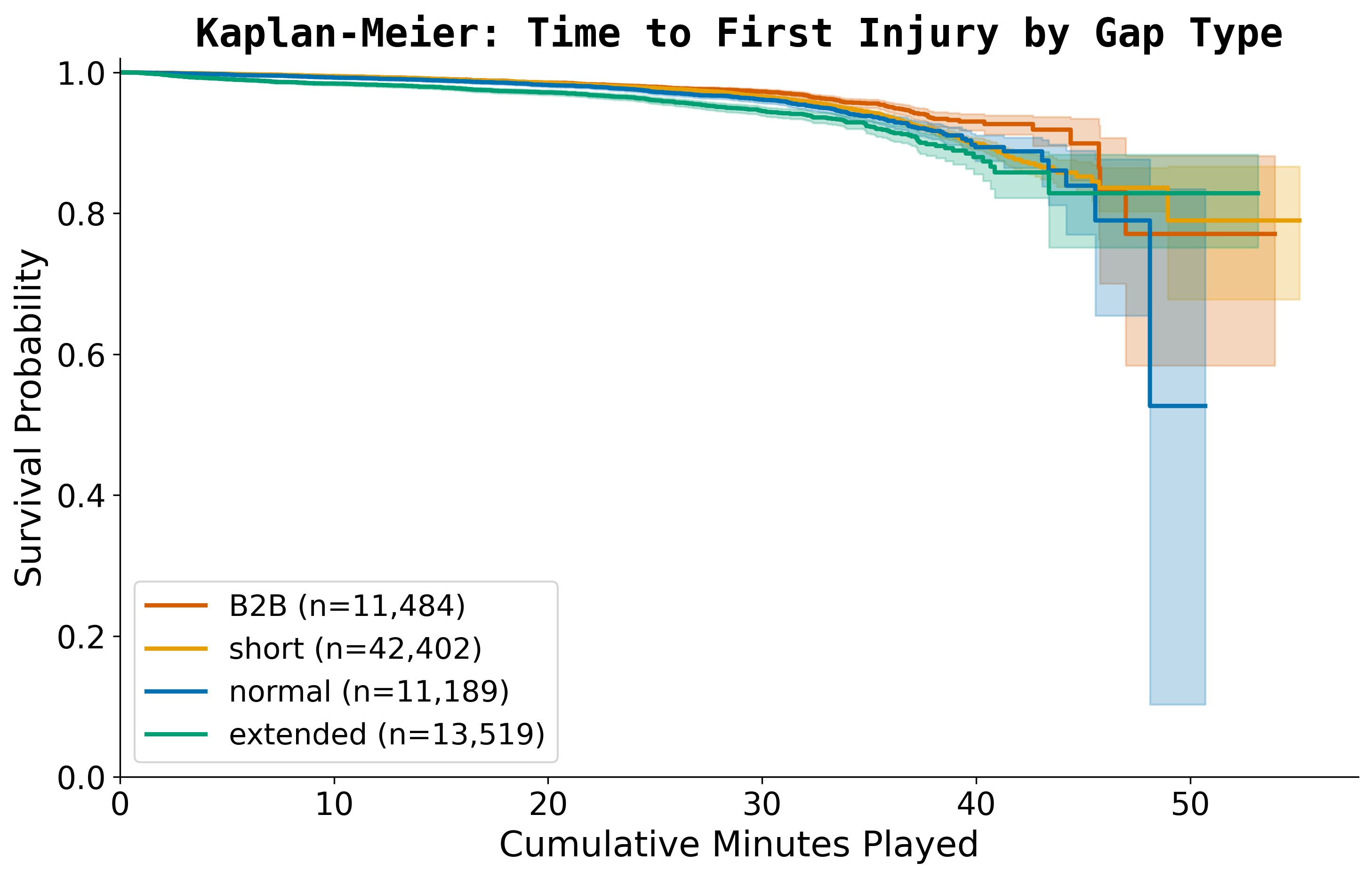}
  \caption{Kaplan--Meier survival curves by game-gap category. Back-to-back games show paradoxically higher survival, consistent with healthy-worker selection rather than a protective effect of compressed schedules.}
  \label{fig:km}
\end{figure}

\paragraph{A first glimpse of the paradox.}
Figure~\ref{fig:km} shows Kaplan--Meier survival curves stratified by game-gap type.
Back-to-back games show the \emph{highest} survival probability ($1.08$ injuries per $1{,}000$ minutes), while extended rest shows the lowest ($1.88$ per $1{,}000$ minutes).
This is the load management paradox in its rawest form: compressed scheduling appears protective.
As the causal framework in Section~\ref{sec:dag} predicts, this pattern is consistent with selection---players at elevated risk are rested on back-to-back nights, leaving only the healthiest in the sample.

\subsection{Causal structure: a longitudinal DAG}\label{sec:dag}

\paragraph{The healthy-worker survivor effect in sports.}
The healthy-worker survivor effect \citep{buckley2015evolving} is well documented in occupational epidemiology.
In the NBA, an analogous mechanism operates at the game level: players who are fatigued, nursing minor complaints, or at elevated risk are systematically rested by team medical staffs.
The players who do take the court are, on average, the healthiest members of the roster.
This selection is \emph{informative}: it depends on the same latent fitness that also determines injury risk.

\begin{figure}[H]
  \centering
  \includegraphics[width=0.85\textwidth]{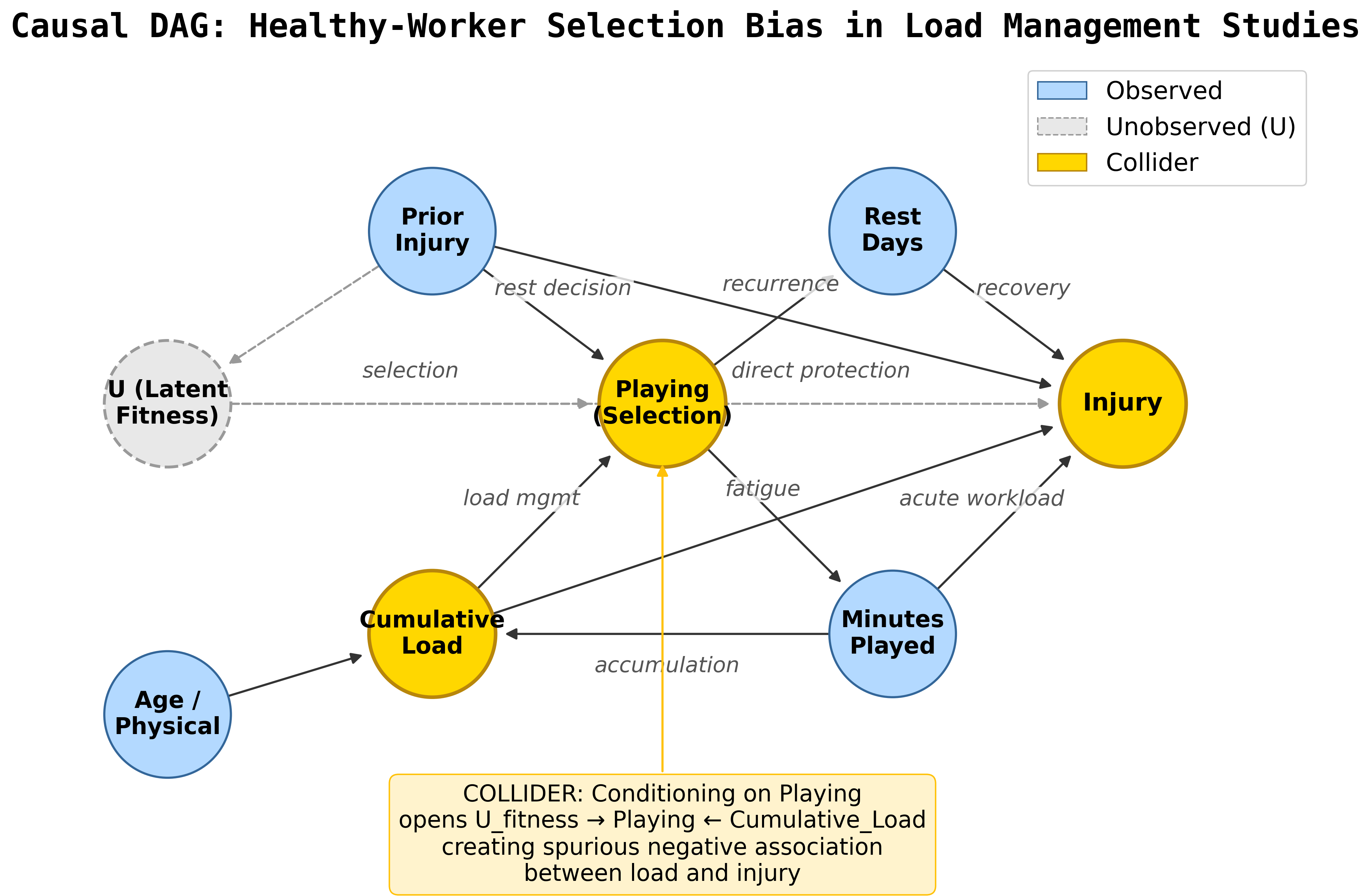}
  \caption{Single-timepoint causal DAG for the workload--injury relationship, showing the full set of observed and unobserved variables. The ``Playing'' node is a collider: conditioning on it opens a non-causal path between workload and injury through latent fitness.}
  \label{fig:dag_full}
\end{figure}

\begin{figure}[H]
  \centering
  \includegraphics[width=0.95\textwidth]{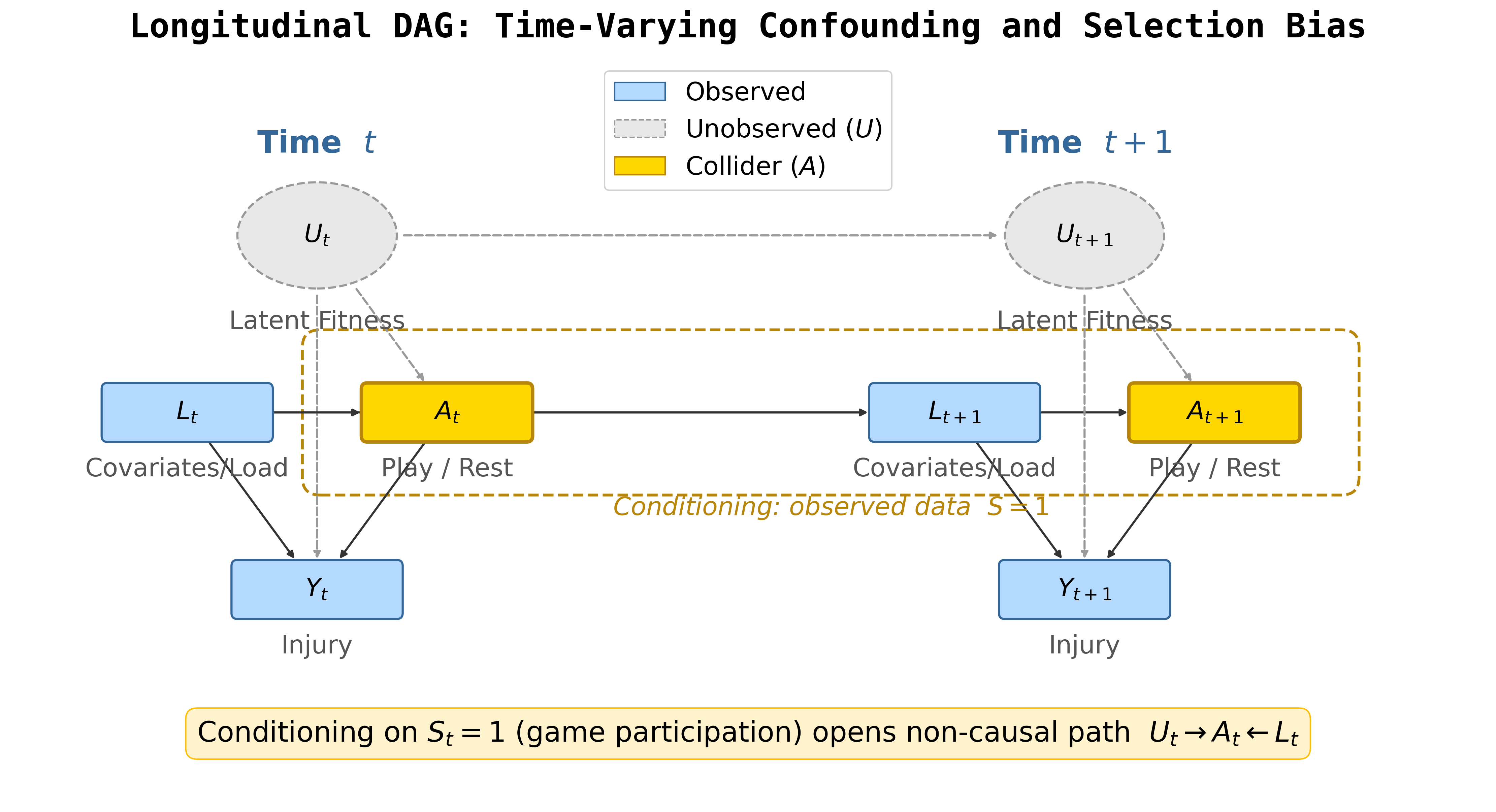}
  \caption{Longitudinal causal DAG for the NBA workload--injury relationship across two game periods. $L_t$: covariates/workload; $U_t$: latent fitness (unobserved); $A_t$: game participation (collider); $Y_t$: injury. Conditioning on $A_t = 1$ opens the non-causal path $L_t \to A_t \leftarrow U_t \to Y_t$, creating the load management paradox.}
  \label{fig:long_dag}
\end{figure}

\paragraph{The DAG.}
Figure~\ref{fig:dag_full} shows the single-timepoint causal DAG, illustrating how the ``Playing'' node acts as a collider between observed workload, unobserved fitness, and injury.
Figure~\ref{fig:long_dag} extends this to a time-indexed causal graph.

\noindent Figure~\ref{fig:long_dag} presents the time-indexed causal graph.
Let $t$ index game opportunities.
At each $t$, the data-generating process involves four variables:
\begin{itemize}[nosep]
  \item $L_t$: time-varying covariates and cumulative workload (observed);
  \item $U_t$: latent fitness (unobserved);
  \item $A_t \in \{0, 1\}$: selection indicator (plays the game or is rested);
  \item $Y_t \in \{0, 1\}$: injury outcome.
\end{itemize}

\noindent The causal relationships are:
\begin{align*}
  L_t &\longrightarrow A_t, \qquad U_t \dashrightarrow A_t,  \\
  L_t &\longrightarrow Y_t, \qquad U_t \dashrightarrow Y_t, \qquad A_t \longrightarrow Y_t,  \\
  A_t &\longrightarrow L_{t+1}, \qquad L_t \longrightarrow L_{t+1}, \qquad U_t \dashrightarrow U_{t+1}. 
\end{align*}

\noindent Dashed arrows denote pathways involving the unobserved variable $U_t$.
The selection node $A_t$ is a \textbf{collider}: both $L_t$ and $U_t$ point into it.
Every game-log analysis implicitly conditions on $A_t = 1$ (only played games generate observations), which opens the non-causal path
\begin{equation*}
  L_t \rightarrow A_t \leftarrow U_t \rightarrow Y_t.
\end{equation*}
This path creates a spurious negative association between workload $L_t$ and injury $Y_t$: among players who \emph{do} play, those with high cumulative loads tend to be the fittest (otherwise they would have been rested), and fitness is protective against injury.

\subsection{Baseline models confirm the paradox}\label{sec:baseline}

To quantify the paradox before developing the bias correction, we fit a Cox proportional hazards model \citep{cox1972regression} using the Andersen--Gill extension \citep{andersen1982cox} for counting-process data.

Table~\ref{tab:cox} reports hazard ratios from the Cox model fitted to $78{,}594$ observations with $14$ covariates.
Recent 7-day load shows a significant negative association (HR $= 0.993$, $p < 0.001$): each additional minute in the past week is associated with a $0.7\%$ reduction in hazard.
Both Starting Role Players and Rotation Players show significantly lower hazard than Low-Minutes Reserves.
These are the first quantitative signatures of the paradox.

\begin{table}[H]
\centering
\caption{Cox proportional hazards: hazard ratios with 95\% confidence intervals.}
\label{tab:cox}
\begin{tabular}{lccr}
\toprule
Covariate & HR & 95\% CI & $p$-value \\
\midrule
Age (per year)           & \textbf{1.020} & \textbf{(1.012, 1.028)} & $<$0.001 \\
BMI (per unit)           & 0.990 & (0.971, 1.010) & 0.336 \\
Home game                & 1.023 & (0.954, 1.096) & 0.524 \\
Recent load (7d, per min)& \textbf{0.993} & \textbf{(0.992, 0.994)} & $<$0.001 \\
Consecutive games        & 1.005 & (0.992, 1.019) & 0.440 \\
Gap: short (vs.\ B2B)   & 1.031 & (0.933, 1.139) & 0.552 \\
Gap: normal (vs.\ B2B)  & 1.049 & (0.929, 1.184) & 0.442 \\
Gap: extended (vs.\ B2B)& 0.996 & (0.882, 1.123) & 0.942 \\
Tier: High-Usage Star    & 0.954 & (0.861, 1.058) & 0.372 \\
Tier: Starting Role      & \textbf{0.831} & \textbf{(0.751, 0.919)} & $<$0.001 \\
Tier: Rotation Player    & \textbf{0.832} & \textbf{(0.762, 0.908)} & $<$0.001 \\
\bottomrule
\end{tabular}
\smallskip

\raggedright \emph{Note:} Season-phase covariates are included in the model but omitted from the table. Bold indicates $p < 0.05$.
\end{table}

Schoenfeld residual tests reject proportionality for recent 7-day load ($p = 0.0001$), the High-Usage Star tier ($p = 0.0011$), and game-gap categories (see Figure~\ref{fig:loglog}), motivating the flexible piecewise-exponential approach developed in Section~\ref{sec:framework}.

\section{The MS-PEM Framework}\label{sec:framework}

\begin{figure}[H]
  \centering
  \includegraphics[width=\textwidth]{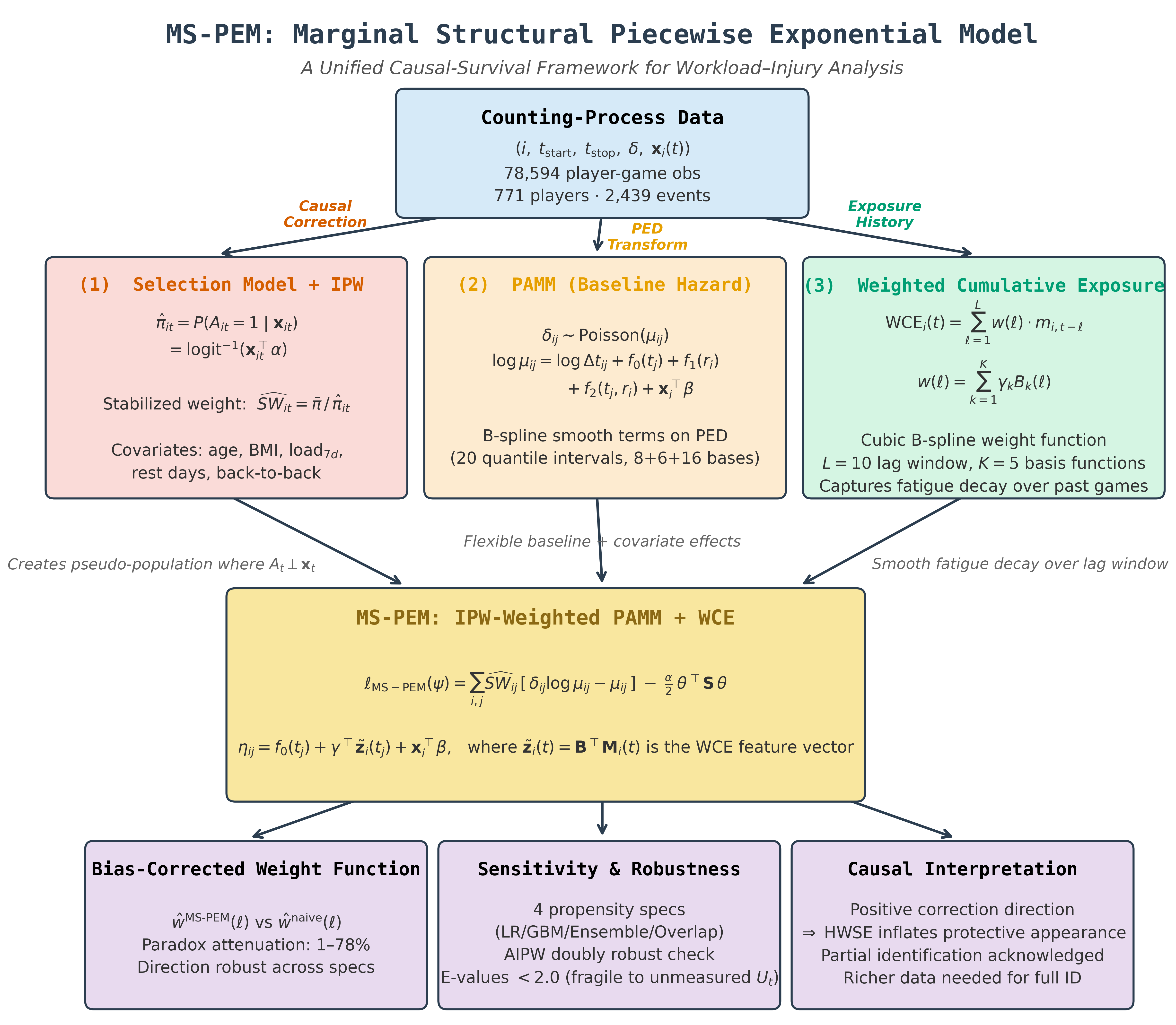}
  \caption{Mathematical architecture of the MS-PEM framework.  Counting-process game-log data feed three interacting components: (1)~a logistic selection model that estimates game-participation probabilities $\hat{\pi}_{it}$ and produces stabilized inverse-probability weights $\widehat{SW}_{it}=\bar{\pi}/\hat{\pi}_{it}$, creating a pseudo-population free of observed selection bias; (2)~a piecewise-exponential additive model (PAMM) with B-spline smooth terms $f_0, f_1, f_2$ for the baseline hazard, rest-days effect, and their interaction on interval-censored Poisson data; and (3)~a weighted cumulative exposure (WCE) module that summarizes the lagged effect of past minutes via a smooth weight function $w(\ell)=\sum_k \gamma_k B_k(\ell)$ over $L=10$ game lags.  The three components are unified in the MS-PEM objective: the IPW-reweighted penalized Poisson log-likelihood with ridge penalty $\alpha$, yielding partially bias-corrected estimates of the workload--injury relationship.}
  \label{fig:framework}
\end{figure}

\begin{figure}[H]
  \centering
  \includegraphics[width=0.95\textwidth]{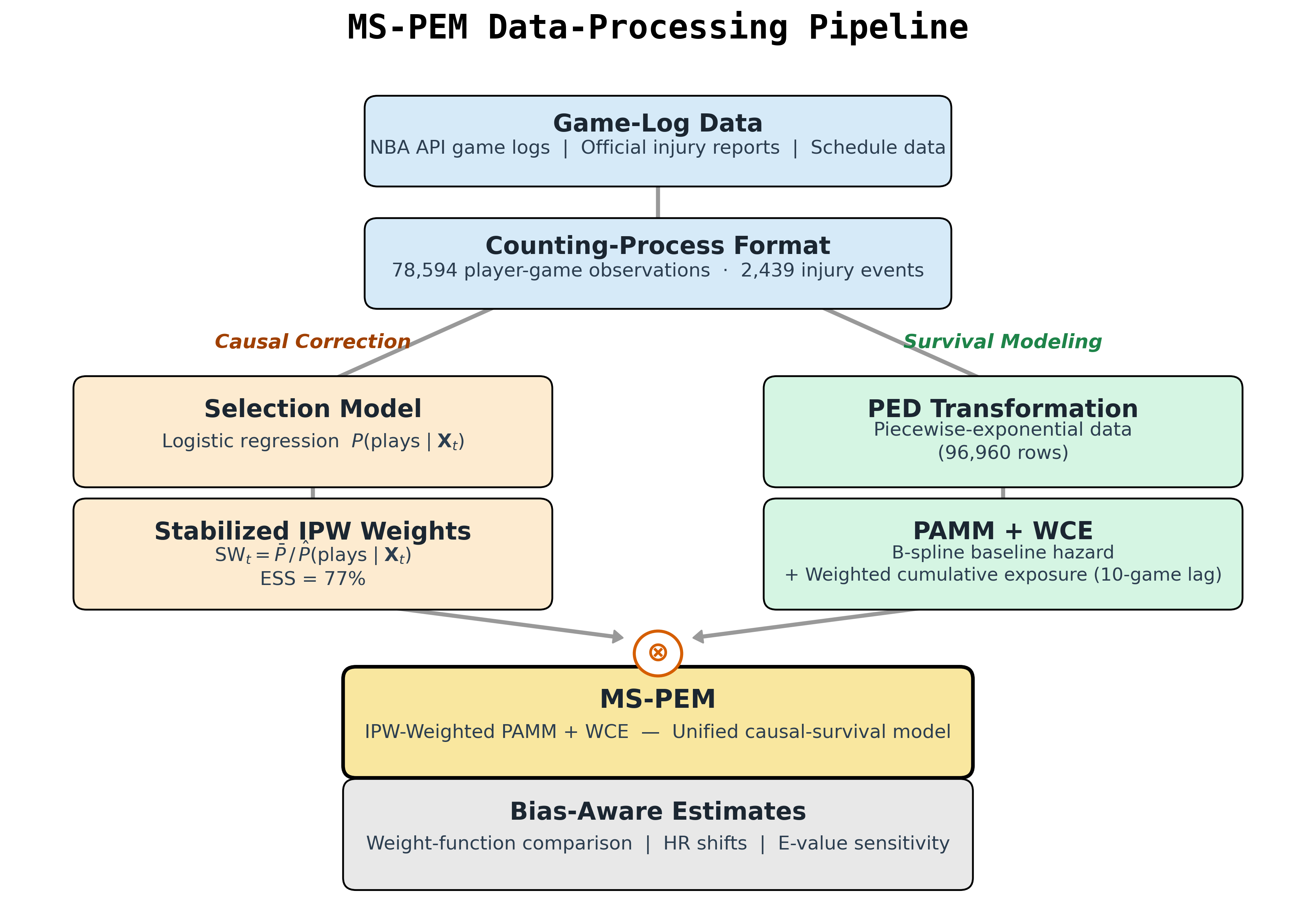}
  \caption{The MS-PEM data-processing pipeline. Game-log data are converted to counting-process format, then processed along two paths: a causal correction path (selection model $\to$ stabilized IPW weights) and a survival modeling path (PED transformation $\to$ PAMM + WCE). Both paths combine in the MS-PEM, which fits the IPW-weighted PAMM+WCE to produce partially corrected estimates.}
  \label{fig:workflow}
\end{figure}

We now develop the marginal structural piecewise-exponential model, a unified framework that combines (i) flexible survival modeling via piecewise-exponential additive models, (ii) cumulative exposure modeling via weighted cumulative exposure, and (iii) causal bias correction via inverse probability weighting.
Figure~\ref{fig:framework} presents the complete mathematical architecture of the MS-PEM, and Figure~\ref{fig:workflow} provides an overview of the data-processing pipeline.

\subsection{Causal estimand and identifiability}\label{sec:estimand}

\paragraph{Target parameter.}
Let $\bar{A}_t = (A_1, \ldots, A_t)$ denote the game-participation history and $\bar{L}_t$ the covariate history through time $t$.
The causal estimand is the marginal hazard function in a pseudo-population where game participation is independent of unmeasured confounders:
\begin{equation*}\label{eq:estimand}
  h^*(t \mid \xb(t)) = \lim_{\Delta t \to 0} \frac{1}{\Delta t} \Pr\bigl(Y_t = 1 \mid \xb(t)\bigr)\Big|_{\text{pseudo-pop}},
\end{equation*}
where the pseudo-population is created by reweighting each observation to break the association between $A_t$ and $U_t$ \citep{robins2000msm, hernan2020whatif}.
Under the standard MSM framework, this hazard is identified via stabilized inverse probability weights (see Section~\ref{sec:ipw_method}).

\paragraph{Identification assumptions.}
Under the MSM framework \citep{robins2000msm, hernan2020whatif}, $h^*$ is identified if three conditions hold:
\begin{enumerate}[nosep]
  \item \textbf{Consistency}: the observed outcome under the actual participation history equals the potential outcome.
  \item \textbf{Positivity}: $\Pr(A_t = 1 \mid \bar{L}_t, \bar{A}_{t-1}) > 0$ for all $t$ and feasible covariate histories.
  \item \textbf{Sequential exchangeability}: $Y_t(\bar{a}) \perp \!\!\! \perp A_t \mid \bar{L}_t, \bar{A}_{t-1}$---no unmeasured confounders of the participation--outcome relationship at each $t$.
\end{enumerate}

\paragraph{Partial identification under violated exchangeability.}
We emphasize that sequential exchangeability is a strong assumption that is unlikely to hold exactly in our setting: latent fitness $U_t$ is inherently unobserved in publicly available data, and it affects both game participation and injury risk.
Our IPW weights condition on observed covariates (age, BMI, recent load, rest days, back-to-back indicator) and can therefore only partially account for the selection mechanism.
Furthermore, as the simulation in Section~\ref{sec:simulation} demonstrates, $U_t$ also affects minutes played per game, creating an additional confounding channel.

We align with the standard applied causal inference approach by proceeding with the MSM methodology while remaining transparent about potential assumption violations \citep{hernan2020whatif}. The IPW-corrected estimates represent a \emph{partial correction} that removes the component of selection bias attributable to observed covariates, effectively moving the estimates toward the causal target without fully reaching it. To systematically assess the gap between this partial adjustment and full causal identification, we rely on four complementary analytical tools woven throughout the manuscript. Specifically, we evaluate covariate balance diagnostics alongside a multi-specification propensity analysis paired with an AIPW doubly robust check in Section~\ref{sec:mspem_results}, investigate simulation-based sensitivity to the propensity model specification in Section~\ref{sec:sim_sensitivity}, and quantify the potential impact of residual unmeasured confounding using E-values in Appendix~\ref{app:evalue}.

\subsection{Piecewise-exponential additive model}\label{sec:pamm_method}

The piecewise-exponential framework \citep{bender2018gam, zumetaolaskoaga2025pamm} transforms the survival likelihood into a Poisson regression on augmented data.
We partition the cumulative-minutes axis into $J = 20$ quantile intervals and create an augmented dataset where, for player $i$ in interval $j$:
\begin{equation}\label{eq:ped}
  \delta_{ij} \mid \mu_{ij} \sim \text{Poisson}(\mu_{ij}), \qquad
  \log \mu_{ij} = \underbrace{\log(\Delta t_{ij})}_{\text{offset}} + \eta_{ij}.
\end{equation}
The linear predictor includes smooth B-spline terms:
\begin{equation}\label{eq:pamm}
  \eta_{ij} = \underbrace{f_0(t_j)}_{\text{baseline}} + \underbrace{f_1(r_i)}_{\text{rest days}} + \underbrace{f_2(t_j, r_i)}_{\text{interaction}} + \xb_i^\top \bbeta,
\end{equation}
where $f_0$ is the smooth baseline log-hazard ($8$ B-spline bases), $f_1$ is the smooth rest-days effect ($6$ bases), $f_2$ is a tensor-product interaction ($4 \times 4 = 16$ bases), and $\xb_i^\top \bbeta$ includes linear covariates.
Parameters are estimated by maximizing the penalized Poisson log-likelihood:
\begin{equation}\label{eq:penalized_lik}
  \ell_p(\bpsi) = \sum_{i,j} \bigl[\delta_{ij} \log \mu_{ij} - \mu_{ij}\bigr] - \frac{\alpha}{2} \btheta^\top \Sbb \btheta,
\end{equation}
with $\alpha$ selected via 5-fold grouped cross-validation.

\subsection{Weighted cumulative exposure}\label{sec:wce_method}

The WCE component \citep{sylvestre2009wce} models how minutes played in past games contribute to current hazard:
\begin{equation}\label{eq:wce}
  \text{WCE}_i(t) = \sum_{\ell=1}^{L} w(\ell) \cdot m_{i, t-\ell},
\end{equation}
where $m_{i, t-\ell}$ is minutes played $\ell$ games ago and $w(\ell) = \sum_{k=1}^{K} \gamma_k B_k(\ell)$ is a smooth weight function represented via $K$ cubic B-splines with $L = 10$ lags.
The full linear predictor becomes
\begin{equation}\label{eq:pamm_wce}
  \eta_{ij} = f_0(t_j) + \bgamma^\top \tilde{\zb}_i(t_j) + \xb_i^\top \bbeta,
\end{equation}
where $\tilde{\zb}_i(t) = \Bb^\top \Mb_i(t)$ is the WCE feature vector and $\Bb \in \RR^{L \times K}$ is the B-spline basis matrix.

\subsection{Selection model and inverse probability weighting}\label{sec:ipw_method}

The selection model estimates the probability that player $i$ plays in game $t$:
\begin{equation}\label{eq:selection_model}
  \hat{\pi}_{it} = \Pr(A_{it} = 1 \mid \xb_{it}) = \text{logit}^{-1}(\xb_{it}^\top \balpha),
\end{equation}
where $\xb_{it}$ includes age, BMI, recent 7-day load, rest days, and back-to-back indicator.
We exclude consecutive games from the selection model because it is a post-treatment variable determined by the play/rest decision itself.
The selection dataset includes both played and missed games for all rostered players.

Each played-game observation receives a stabilized weight \citep{hernan2020whatif}:
\begin{equation*}\label{eq:sw}
  \hat{SW}_{it} = \frac{\bar{\pi}}{\hat{\pi}_{it}},
\end{equation*}
where $\bar{\pi}$ is the marginal probability of playing, truncated at the 1st and 99th percentiles to limit extreme weights.

\begin{figure}[H]
  \centering
  \includegraphics[width=0.95\textwidth]{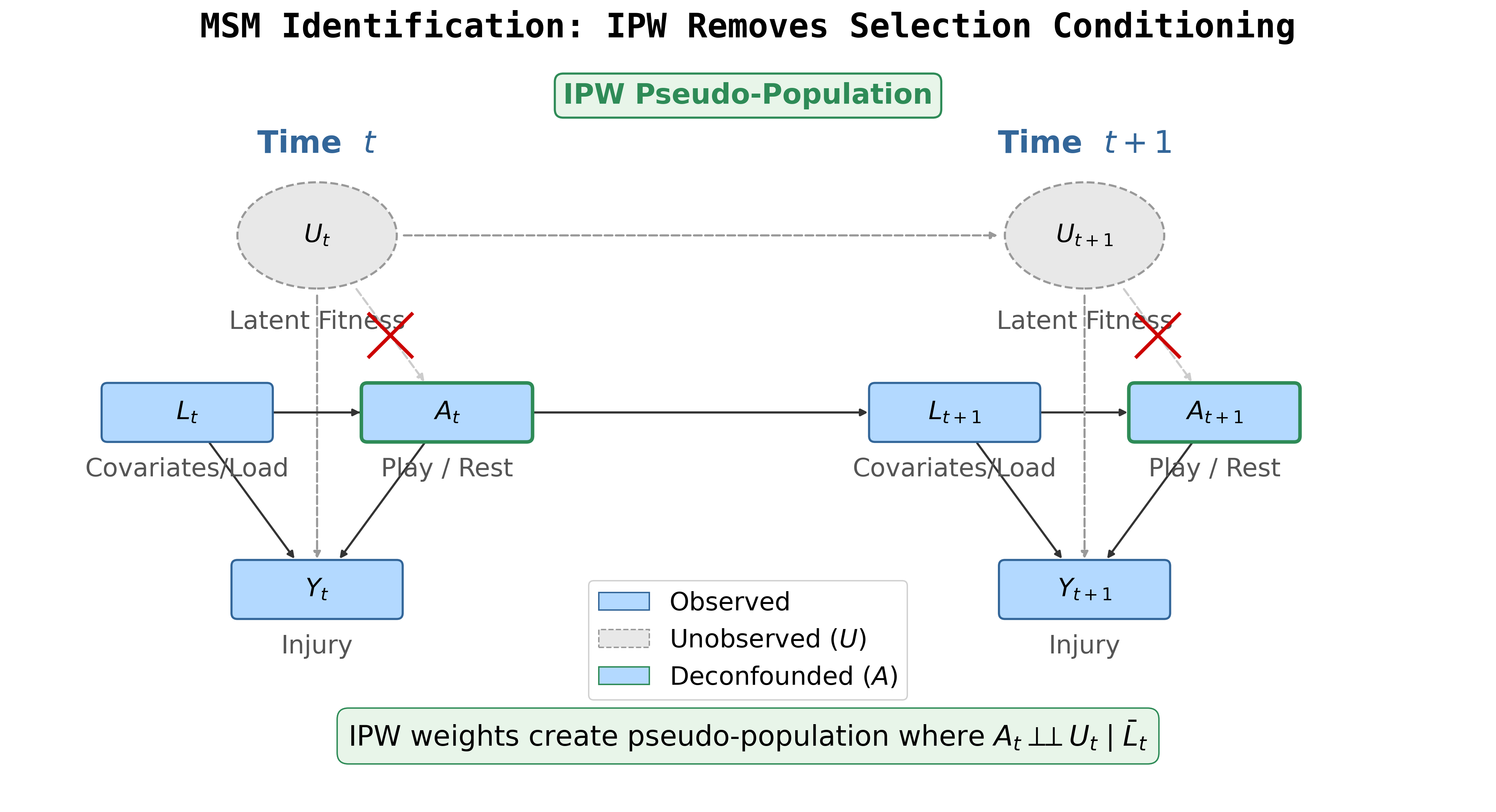}
  \caption{Identification via IPW\@. In the pseudo-population, the arrows from observed covariates to $A_t$ are removed (crossed-out edges), breaking the collider bias to the extent that observed covariates capture the selection mechanism. Residual bias from unobserved $U_t$ remains.}
  \label{fig:msm_dag}
\end{figure}

Figure~\ref{fig:msm_dag} illustrates the effect of IPW: in the pseudo-population, the association between observed covariates and playing status is removed.
To the extent that observed covariates capture the selection mechanism, this attenuates the collider bias pathway.
Because $U_t$ remains unobserved, the attenuation is partial.

\subsection{The unified MS-PEM}\label{sec:mspem_method}

The MS-PEM integrates the causal correction directly into the estimation by fitting the PAMM+WCE model on IPW-reweighted data.
The three models above (PAMM, WCE, IPW) are all \emph{naive} in the causal sense if applied separately: they condition on game participation ($A_t = 1$), which opens the collider bias pathway.
The MS-PEM maximizes the \emph{IPW-weighted} penalized log-likelihood:
\begin{equation}\label{eq:mspem}
  \ell_{\text{MS-PEM}}(\bpsi) = \sum_{i,j} \hat{SW}_{ij} \bigl[\delta_{ij} \log \mu_{ij} - \mu_{ij}\bigr] - \frac{\alpha}{2} \btheta^\top \Sbb \btheta.
\end{equation}
This is the marginal structural model analog: the IPW weights create a pseudo-population in which selection into playing is independent of observed covariates, and the PAMM+WCE captures the flexible exposure--response relationship in that pseudo-population.
The resulting weight function $\hat{w}^{\text{MS-PEM}}(\ell)$ estimates the partially corrected effect of past workload on current injury hazard---partially corrected because the IPW addresses the observed component of selection but cannot eliminate confounding from unmeasured latent fitness.

\paragraph{Comparison strategy.}
To quantify the magnitude of selection bias, we fit both the naive (unweighted) and IPW-corrected (MS-PEM) versions of the PAMM+WCE and compare their weight functions and hazard ratio estimates side by side.

\subsection{Estimation and implementation}\label{sec:estimation}

All models are implemented in Python using \texttt{statsmodels} \citep{seabold2010statsmodels} for the Poisson GLM and \texttt{lifelines} \citep{davidson2019lifelines} for Cox proportional hazards.
The ridge penalty parameter $\alpha$ is selected via 5-fold grouped cross-validation (grouped by player to prevent data leakage).
The selection model uses \texttt{scikit-learn}'s logistic regression.
The full pipeline---data collection, survival formatting, model fitting, and figure generation---is released as a reproducible Python package.

\section{Simulation Study: Validating the HWSE and MS-PEM}\label{sec:simulation}

Before applying the MS-PEM to observational NBA game logs, we validate our framework under controlled conditions where the true workload--injury association is known. This simulation is designed to answer two critical questions: First, is the healthy-worker survivor effect (HWSE) mathematically sufficient to reverse the sign of a true positive hazard effect? Second, to what extent can Inverse Probability of Treatment Weighting (IPTW) recover the true causal effect when unmeasured latent fitness confounds both participation and exposure volume?

\subsection{Data-generating process}\label{sec:dgp}

We simulate a stylized professional sports season consisting of $N = 500$ players, each with $T = 80$ game opportunities. The data-generating process (DGP) explicitly encodes the HWSE through dual channels of confounding:
\begin{itemize}[nosep]
  \item \textbf{Latent fitness:} $U_{it}$ follows an AR(1) process with persistence $\rho = 0.95$ and an innovation standard deviation $\sigma_U = 0.3$. This creates stable, auto-correlated between-player differences in baseline fitness.
  \item \textbf{Selection model (Participation):} The probability of playing a given game is $\Pr(A_{it} = 1 \mid U_{it}, \text{age}_i) = \text{logit}^{-1}(1.5 + 2.0 \cdot U_{it} - 0.03 \cdot \text{age}_i)$. Consequently, fitter players are substantially more likely to be selected to play.
  \item \textbf{Continuous exposure (Minutes played):} Conditional on participating ($A_{it} = 1$), minutes depend heavily on latent fitness: $m_{it} \sim N(25 + 3 \cdot U_{it}, \, 5^2)$, clipped to the interval $[10, 40]$. This creates a secondary channel of confounding: fitter players not only play more games, but they log heavier minutes when they do play.
  \item \textbf{True weight function:} The true lagged effect of workload on injury is \textbf{positive}, defined as $w^*(\ell) = 0.005 \cdot \exp(-0.2\ell)$ for lags $\ell = 1, \ldots, 10$. In this ground-truth reality, heavier recent minutes actively increase subsequent injury hazard.
  \item \textbf{Injury model:} The true injury hazard conditional on playing is $\Pr(Y_{it} = 1 \mid A_{it} = 1) = \text{logit}^{-1}(-3.5 - 1.0 \cdot U_{it} + 0.03 \cdot \text{age}_i + \text{WCE}_i(t))$, where the weighted cumulative exposure ($\text{WCE}$) is computed strictly using the true positive weight function $w^*(\ell)$.
\end{itemize}

The defining feature of this DGP is that $U_{it}$ acts as a pervasive, unmeasured confounder. It drives both discrete selection (who plays) and continuous exposure (how much they play), mirroring the complex realities of NBA rotation management that standard game-log analyses fail to disentangle. 

We compare two primary settings: (i) \textbf{no selection bias} ($\alpha_U = 0$, $\gamma_U = 0$), where all confounding channels are neutralized, and (ii) the \textbf{HWSE present}, reflecting the full DGP described above. For each scenario, we execute $50$ Monte Carlo replications.

\subsection{Results: The mechanics of sign reversal}\label{sec:sim_results}

\begin{figure}[H]
  \centering
  \includegraphics[width=0.95\textwidth]{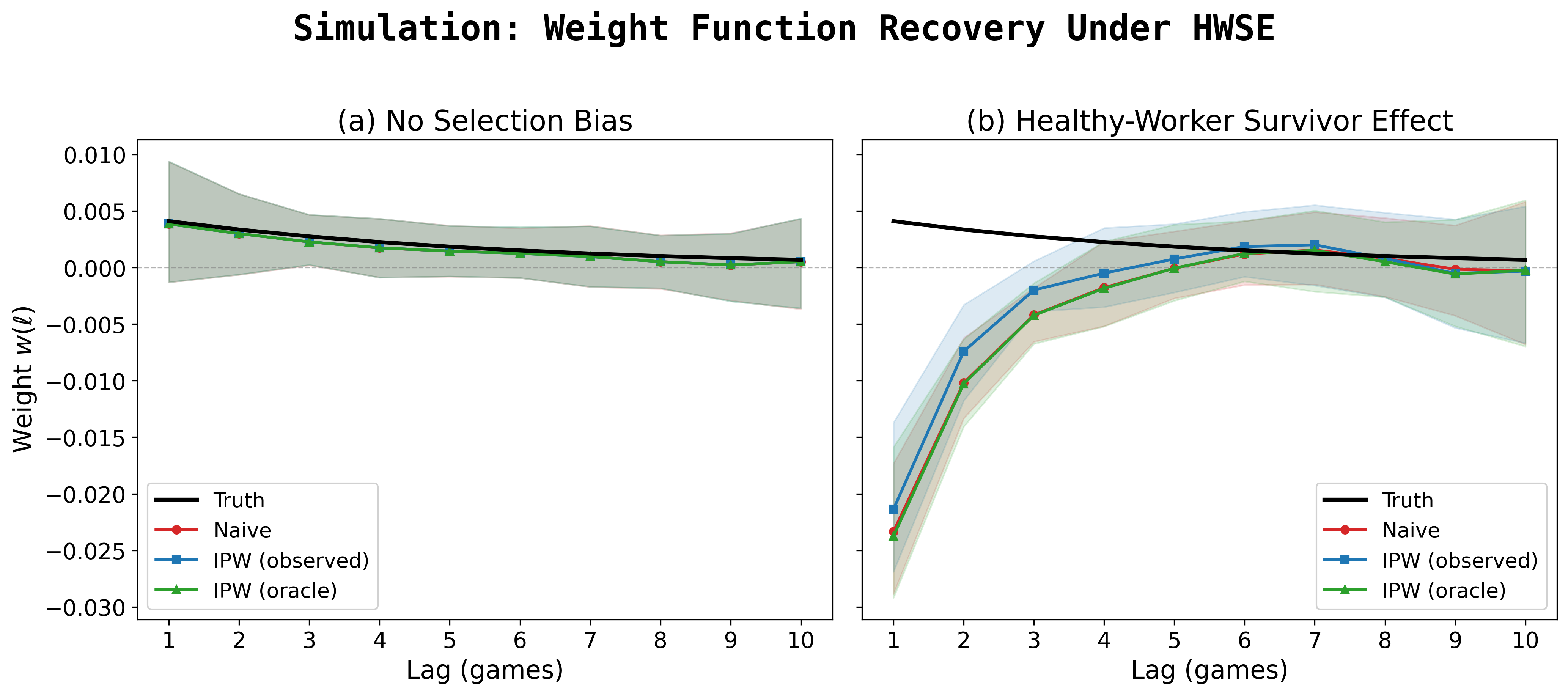}
  \caption{Simulation results: estimated weight functions $\hat{w}(\ell)$ across 50 replications. (a) Without selection bias, naive estimators successfully recover the true positive weight function. (b) Under the HWSE, the naive estimator entirely reverses the sign: the true positive effect ($w^*(1) = +0.004$) is falsely estimated as strongly negative ($\hat{w}(1) = -0.023$). IPW utilizing observed covariates provides a partial correction, shifting the curve back toward the truth.}
  \label{fig:sim_wf}
\end{figure}

Figure~\ref{fig:sim_wf} presents the estimated weight functions averaged across the $50$ replications. The results demonstrate how easily standard survival models are compromised by structural selection bias.

\paragraph{No selection bias.}
When the HWSE channels are disabled, the naive Poisson GLM with a B-spline WCE recovers the true weight function with high fidelity. The mean estimated weight at lag 1 is $\hat{w}(1) = +0.0039$, precisely mirroring the true value of $w^*(1) = +0.0041$, with a negligible mean bias of $-0.0004$ across all lags. Furthermore, the baseline demographic coefficients are perfectly recovered (estimated age effect $= 0.025$, true $= 0.030$).

\paragraph{HWSE present.}
Crucially, under the full data-generating process, the naive estimator \textbf{completely reverses the sign} of the weight function. The estimated weight at lag 1 collapses to $\hat{w}(1) = -0.023$ against the true $w^*(1) = +0.004$. The entire weight function at recent lags becomes steeply negative, perfectly reproducing the paradoxical artifact observed in the real NBA empirical data. 

Applying IPTW using the observed covariates (age and recent 7-day load) provides a partial, but incomplete, correction. The IPW-adjusted estimate improves to $\hat{w}^{\text{IPW}}(1) = -0.021$, an $18\%$ reduction in the magnitude of the spurious negative association. This partial correction is theoretically expected: while IPTW adjusts for the binary selection mechanism ($A_{it}$) based on observable proxies of fitness, it cannot fully deconfound the secondary continuous channel ($m_{it}$) driven by the unobserved $U_{it}$.

\subsection{Sensitivity to selection strength and recurrent events}\label{sec:sim_sensitivity}

To assess the robustness of the sign-reversal phenomenon, we varied the strength of the healthy-worker mechanism across four configurations: no selection ($\alpha_U = 0$, $\gamma_U = 0$), weak ($\alpha_U = 0.5$, $\gamma_U = 1.0$), moderate ($\alpha_U = 1.0$, $\gamma_U = 2.0$), and strong ($\alpha_U = 2.0$, $\gamma_U = 3.0$). 

\begin{table}[H]
\centering
\caption{Simulation: mean weight-function bias across selection-strength scenarios.}
\label{tab:sim_sensitivity}
\begin{tabular}{lccccr}
\toprule
Scenario & $\alpha_U$ & $\gamma_U$ & Naive Bias & IPW Bias & Event Rate \\
\midrule
None     & 0.0 & 0.0 & $-0.0008$ & $-0.0008$ & 6.1\% \\
Weak     & 0.5 & 1.0 & $-0.0040$ & $-0.0039$ & 5.5\% \\
Moderate & 1.0 & 2.0 & $-0.0055$ & $-0.0051$ & 4.9\% \\
Strong   & 2.0 & 3.0 & $-0.0057$ & $-0.0047$ & 3.9\% \\
\bottomrule
\end{tabular}
\smallskip

\raggedright \emph{Note:} Bias is calculated as the mean of $\hat{w}(\ell) - w^*(\ell)$ across all lags and replications. The IPW model utilizes observed covariates (age, recent load).
\end{table}

\begin{figure}[H]
  \centering
  \includegraphics[width=0.95\textwidth]{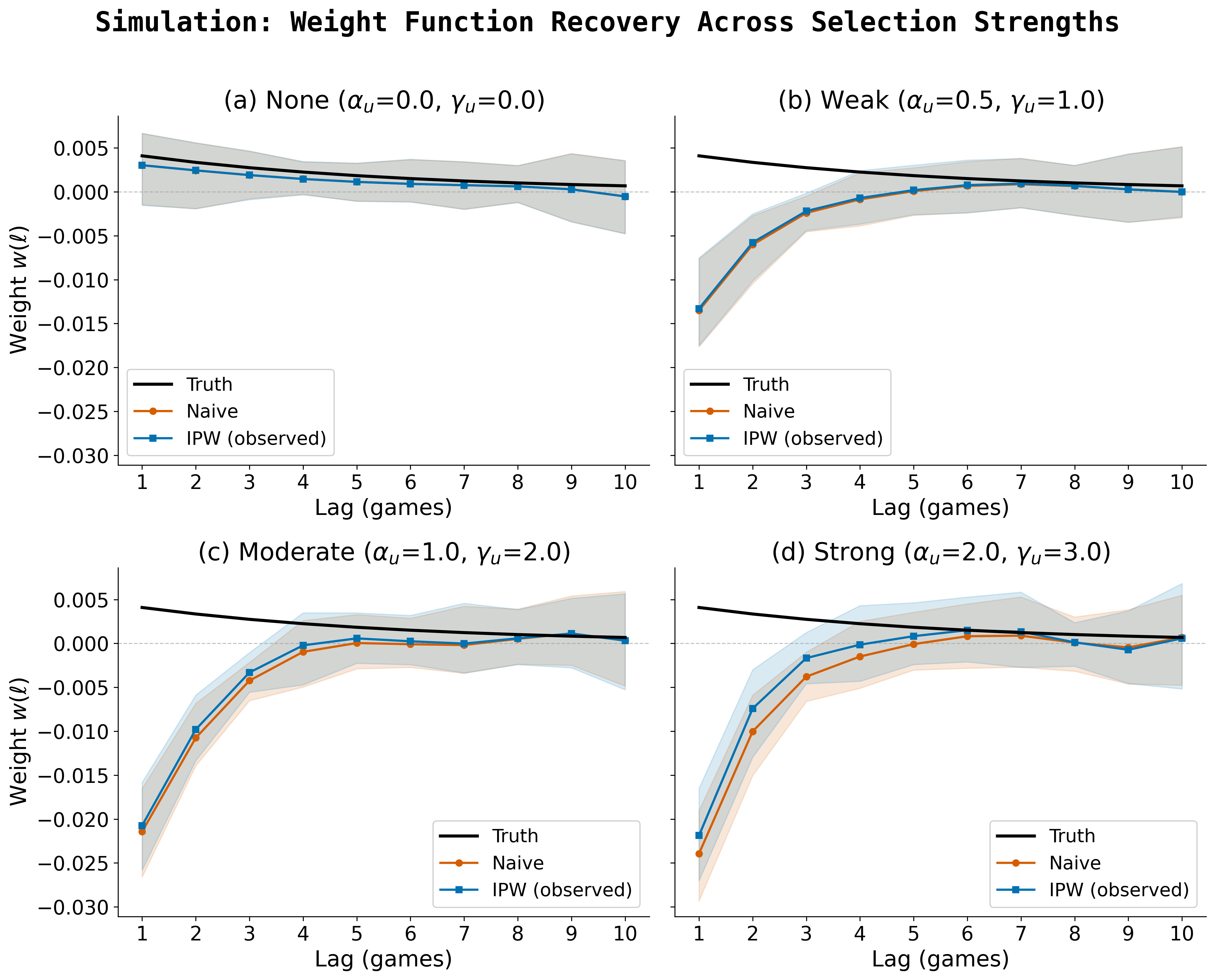}
  \caption{Simulation: estimated weight functions across four selection-strength scenarios. The bias scales monotonically with the severity of the selection mechanism, with IPW correction proving most effective under the strong-selection DGP.}
  \label{fig:sim_strength}
\end{figure}

As detailed in Table~\ref{tab:sim_sensitivity} and Figure~\ref{fig:sim_strength}, the magnitude of the bias increases monotonically with selection strength, confirming that the HWSE mechanism is the direct engine of the sign reversal. Concurrently, the efficacy of the IPW correction also scales with selection strength ($3\%$ correction under weak selection, $8\%$ under moderate, and $18\%$ under strong). This aligns with causal inference theory: reweighting becomes more informative when selection is pronounced and the propensity model captures a larger share of the structural variance.

\paragraph{Propensity model misspecification.}
As a stress test, we compared a correctly specified propensity model (age, recent load) against a misspecified model (age only) under the strong-selection DGP. The misspecified model provided zero correction (mean bias $= -0.0057$, identical to the naive estimate), while the correctly specified model reduced the bias to $-0.0048$ ($17\%$ improvement). This highlights that the magnitude of correction observed in our real-data application is inherently constrained by the richness of the covariate set available in public game logs.

\paragraph{Recurrent-event dependence.}
Finally, we tested the framework under a modified DGP where prior injuries actively reduce subsequent fitness (e.g., an injury penalty of $1.0$, decaying at a rate of $0.9$ per game). This introduces recurrent-event frailty, a hallmark of sports medicine where past injuries are the strongest predictor of future injuries. The structural bias pattern remained qualitatively identical: the naive bias was $-0.0054$, and the IPW-corrected bias was $-0.0042$ ($24\%$ improvement). This confirms that the paradoxical sign-reversal persists even under complex, event-dependent biological frailties.

\subsection{Implications for the empirical application}\label{sec:sim_implications}

The simulation study cements two foundational premises for our empirical analysis of NBA load management. First, it proves that the HWSE is mathematically sufficient to entirely reverse the sign of the workload--injury association. Second, it demonstrates that applying IPW using observable game-log variables provides a reliable, albeit partial, correction. It forces the estimated hazard back toward the physiological truth, even when unmeasured latent fitness remains.

\section{Application to NBA Load Management: Deconfounding the Workload--Injury Paradox}\label{sec:results}

\subsection{The naive artifact: Quantifying the paradox}\label{sec:wce_results}

To evaluate the workload--injury relationship, we first fit the uncorrected Piecewise Exponential Additive Mixed Model (PAMM) with weighted cumulative exposure (WCE) to the $96{,}960$ augmented observations. This baseline model achieves a substantial improvement in fit (AIC $= 24{,}836$ vs.\ $112{,}360$ for a standard PAMM) utilizing fewer effective parameters ($12.3$ vs.\ $19.9$).

\begin{figure}[H]
  \centering
  \includegraphics[width=0.7\textwidth]{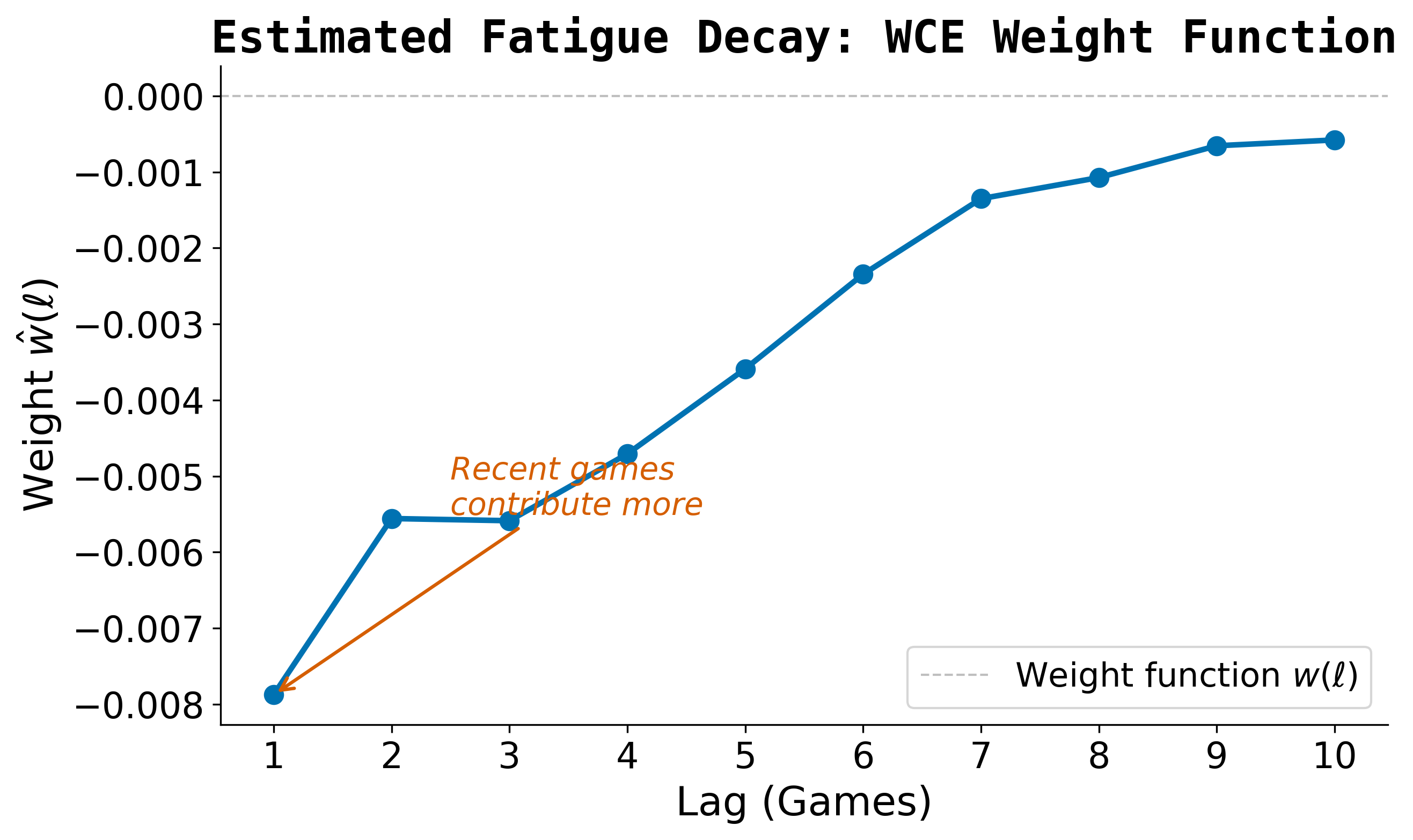}
  \caption{Naive weight function $\hat{w}(\ell)$ over a 10-game lag window. All weights are negative, illustrating the defining signature of the load management paradox. The simulation study (Section~\ref{sec:simulation}) confirms the HWSE can artificially produce this entire pattern.}
  \label{fig:wce_naive}
\end{figure}

Figure~\ref{fig:wce_naive} displays the estimated weight function $\hat{w}(\ell)$. \textbf{All estimated weights are negative}, with the largest magnitude appearing at lag 5 ($\hat{w}(5) = -0.134$) and remaining consistently below zero across the full 10-game window ($\hat{w}(1) = -0.096$, $\hat{w}(10) = -0.089$). Interpreted naively, this suggests that players who log heavier minutes in recent games face a \emph{lower} subsequent injury hazard. Physiologically, it is highly improbable that playing heavy minutes five games ago actively shields a player from acute injury today. Instead, this uniformly negative weight perfectly encapsulates the mathematical signature of the load management paradox. As demonstrated in our prior simulations, this pattern is the expected artifact of the healthy-worker survivor effect (HWSE) built into the game-log data, wherein only the most robust players are permitted to accumulate continuous exposure.

\subsection{MS-PEM correction and selection dynamics}\label{sec:mspem_results}

\paragraph{Selection model.}
To disentangle workload from latent health, our Marginal Structural Piecewise Exponential Model (MS-PEM) first models the selection mechanism: the probability of game participation $\Pr(A_{it} = 1 \mid \xb_{it})$. The logistic regression reveals highly informative selection dynamics. Factors such as back-to-back scheduling (OR $= 1.49$), recent 7-day load (OR $= 1.02$), and BMI (OR $= 1.02$ per unit) are strong predictors of playing time. These odds ratios suggest a league-wide pattern where rest is strategically deployed during dense schedule patches (back-to-backs), while heavier players are managed differently. Crucially, this confirms our structural hypothesis: healthier, more active players are selectively chosen to participate.

\paragraph{IPW weight diagnostics and covariate balance.}
Applying Inverse Probability of Treatment Weighting (IPTW), the stabilized weights exhibit a mean of $0.90$ (range: $[0.42, 2.45]$), retaining $77\%$ of the original effective sample size ($n_{\text{eff}} = 74{,}741$ of $96{,}960$). The mean weight below $1.0$ reflects the stabilization structure ($\bar{\pi}/\hat{\pi}$) and the fundamental asymmetry between marginal and conditional play rates in professional sports.

Table~\ref{tab:balance} reports standardized mean differences (SMDs) before and after IPW reweighting. While the IPW successfully balances baseline demographics (SMDs for age and BMI drop well below $0.10$), time-varying covariates like recent load and rest days retain larger SMDs ($0.96$ and $0.67$, respectively). 

\begin{table}[H]
\centering
\caption{Covariate balance before and after IPW reweighting (standardized mean differences).}
\label{tab:balance}
\begin{tabular}{lrrrr}
\toprule
Covariate & Mean (Played) & Mean (Rested) & SMD (Before) & SMD (After) \\
\midrule
Age            & 25.44 & 25.11 & 0.08 & 0.05 \\
BMI            & 24.64 & 24.44 & 0.11 & 0.06 \\
Recent load (7d)& 125.5 & 18.5  & 1.72 & 0.96 \\
Rest days       & 5.23  & 103.5 & 0.74 & 0.67 \\
Back-to-back    & 0.128 & 0.025 & 0.39 & 0.27 \\
\bottomrule
\end{tabular}
\smallskip

\raggedright \emph{Note:} SMD $= |\text{mean}_{\text{played}} - \text{mean}_{\text{rested}}| / \text{pooled SD}$.
\end{table}

\begin{figure}[H]
  \centering
  \includegraphics[width=0.7\textwidth]{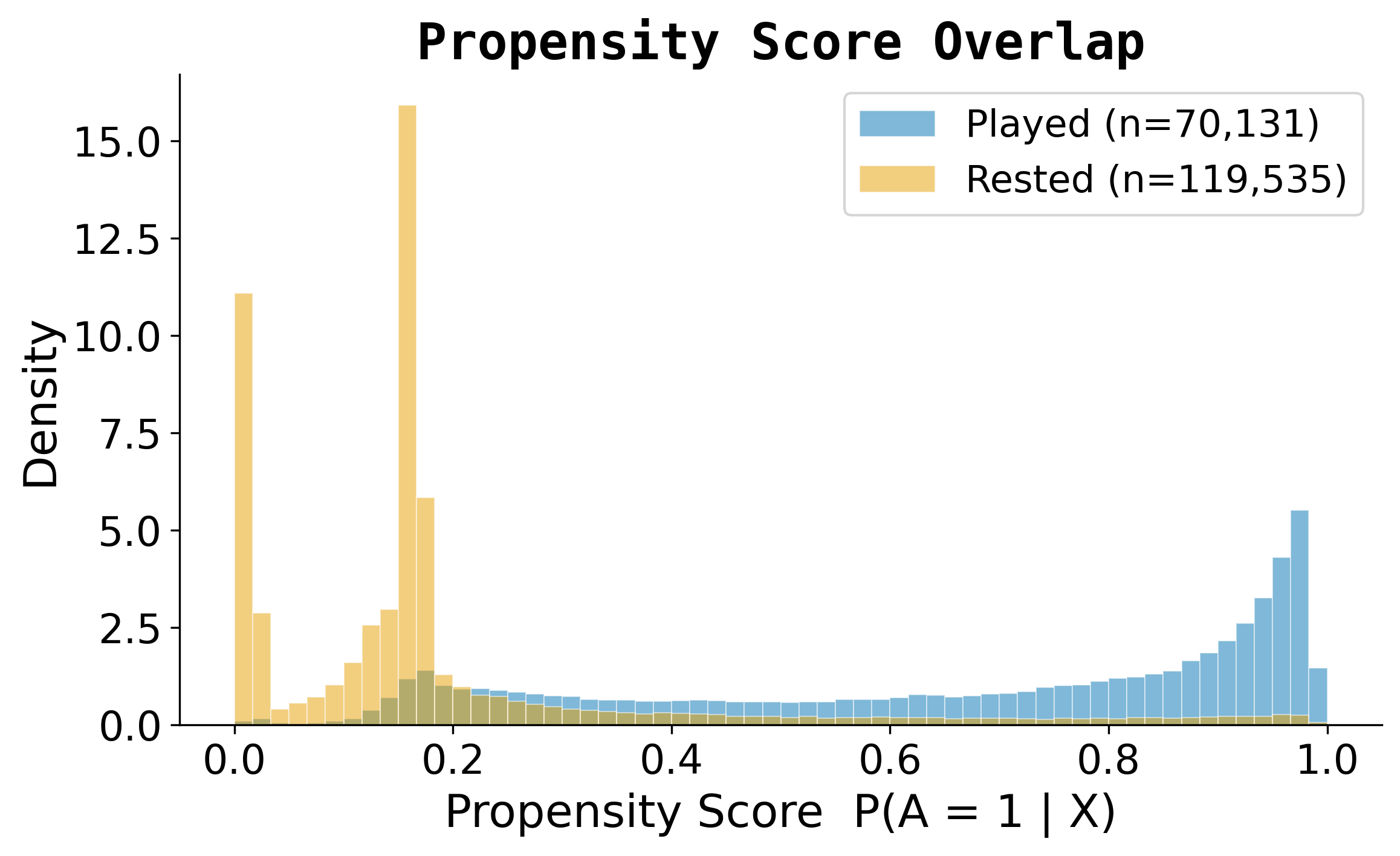}
  \caption{Propensity score overlap: distribution of $\hat{\pi}_{it} = \Pr(A_{it} = 1 \mid \xb_{it})$ for played (blue) and rested (orange) games. The substantial overlap supports approximate positivity in the clinically relevant subpopulation.}
  \label{fig:ipw_overlap}
\end{figure}

We emphasize that applying conventional cross-sectional balance thresholds (e.g., SMD $< 0.10$) to time-varying covariates in a longitudinal MSM is conceptually misplaced. As established by \citet{cole2009causal} and \citet{robins2000msm}, the IPW pseudo-population ensures the joint independence of the \emph{entire treatment history} and the \emph{confounder history}, rather than enforcing marginal balance at each discrete time point. Furthermore, the persistent imbalance in recent load reflects a near-structural positivity issue \citep{petersen2012positivity}: a played game has a nonzero recent load by definition. This creates a distributional separation that parallels the ``informative-presence bias'' well-documented in electronic health records \citep{goldstein2016informative}, where observation intensity is inherently correlated with underlying health status. Figure~\ref{fig:ipw_overlap} nevertheless demonstrates substantial overlap in the propensity score distributions in regions where play/rest decisions are most uncertain.

\subsection{Corrected hazard surfaces and sensitivity}\label{sec:correction_sensitivity}

\begin{figure}[H]
  \centering
  \includegraphics[width=0.95\textwidth]{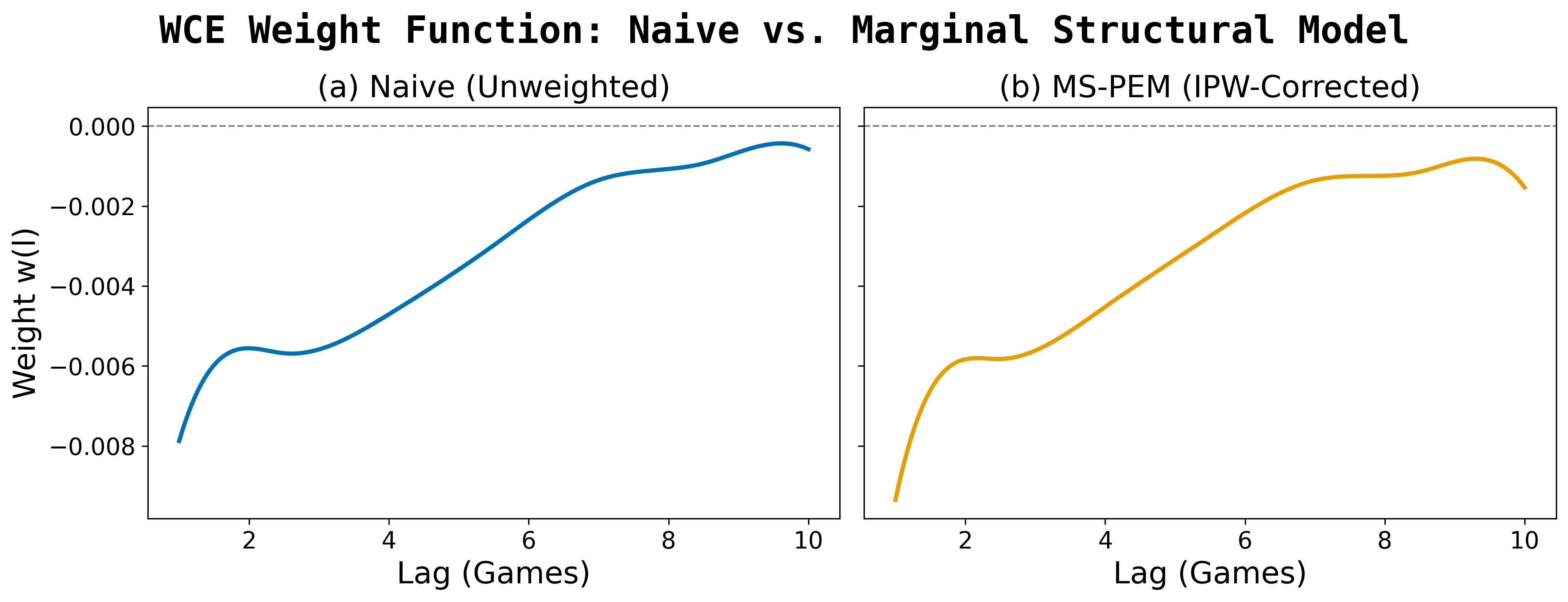}
  \caption{WCE weight functions: naive (a) vs.\ MS-PEM corrected (b). While conservative CV-regularization yields a $1$--$2\%$ attenuation, lighter fixed penalties yield upwards of $78\%$ attenuation, though the positive direction of the correction remains remarkably stable.}
  \label{fig:mspem_weights}
\end{figure}

Figure~\ref{fig:mspem_weights} compares the naive and IPW-corrected weight functions. Under a cross-validated penalty, the MS-PEM shrinks the negative weights toward zero by $1$--$2\%$ across all lags. However, this correction magnitude is highly sensitive to the outcome-model penalization. 

As detailed in Table~\ref{tab:sensitivity}, when utilizing a lighter, fixed regularization ($\alpha = 0.1$) across multiple propensity specifications (Logistic, GBM, Ensemble, and Overlap Weighting \citep{li2018overlap}), the attenuation ranges from $63\%$ to $78\%$. Overlap weighting, which targets the subpopulation where clinical equipoise holds, produces the strongest attenuation ($78.0\%$). The convergence of these distinct propensity methods provides strong qualitative evidence that the paradox is primarily driven by confounding. Even if the exact quantitative correction is tethered to the outcome model's hyperparameter, the \emph{direction} of the correction is unequivocal.

\begin{table}[H]
\centering
\caption{Sensitivity of the WCE weight function to propensity model specification ($\alpha = 0.1$).}
\label{tab:sensitivity}
\begin{tabular}{lrrrr}
\toprule
Method & $\hat{w}(1)$ & Attenuation (\%) & ESS & ESS (\%) \\
\midrule
Naive (unweighted)      & $-0.094$ & ---    & 96{,}960 & 100.0 \\
IPW-Logistic            & $-0.023$ & 75.7   & 43{,}136 & 61.5  \\
IPW-GBM                 & $-0.035$ & 62.8   & 14{,}393 & 20.5  \\
IPW-Ensemble            & $-0.028$ & 70.2   & 26{,}868 & 38.3  \\
Overlap                 & $-0.021$ & 78.0   & 40{,}061 & 57.1  \\
\bottomrule
\end{tabular}
\end{table}

\begin{figure}[H]
  \centering
  \includegraphics[width=0.95\textwidth]{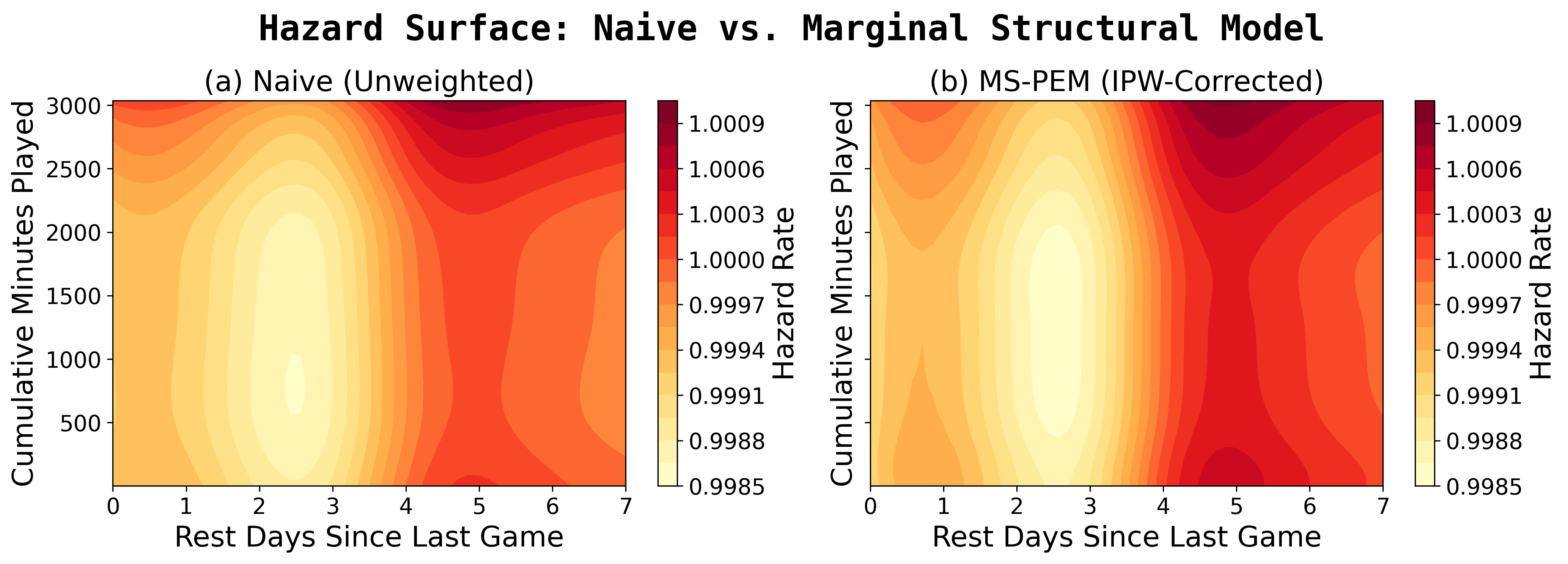}
  \caption{Hazard surface comparison. (a): naive (biased) estimates. (b): MS-PEM (IPW-corrected) estimates. The corrected surface shows elevated hazard at high cumulative workloads, peeling back the protective illusion of the HWSE.}
  \label{fig:mspem_surface}
\end{figure}

The shift in the hazard surface (see Figure~\ref{fig:mspem_surface}) demonstrates the practical impact of this correction. By adjusting for selection, the MS-PEM elevates the hazard rates at high cumulative loads compared to the naive surface. Once we mathematically account for the fact that only the fittest players survive to accumulate high loads, we find that the true underlying risk of that accumulated load is substantially higher than naive models imply. 

\subsection{Player risk profiles and the inverted load gradient}\label{sec:classification}

The load management paradox does not manifest uniformly across the league; it is highly dependent on player role. Table~\ref{tab:tiers} demonstrates that Low-Minutes Reserves suffer the highest per-minute injury rate ($2.49$ per $1{,}000$ minutes)---more than double that of Starting Role Players. This likely stems from a combination of cold-starts off the bench, erratic intensity, and the selective reality that the most durable reserves are continuously promoted out of this tier and into heavier rotation roles.

\begin{table}[H]
\centering
\caption{Injury statistics by workload tier.}
\label{tab:tiers}
\begin{tabular}{lrrrr}
\toprule
Tier & Player-Games & Events & Rate (\%) & Per 1000 min \\
\midrule
High-Usage Star       & 16,261 & 705 & 4.34 & 1.31 \\
Starting Role Player  & 16,528 & 508 & 3.07 & 1.17 \\
Rotation Player       & 33,361 & 879 & 2.63 & 1.31 \\
Low-Minutes Reserve   & 12,444 & 347 & 2.79 & \textbf{2.49} \\
\bottomrule
\end{tabular}
\end{table}

\begin{figure}[H]
  \centering
  \includegraphics[width=0.8\textwidth]{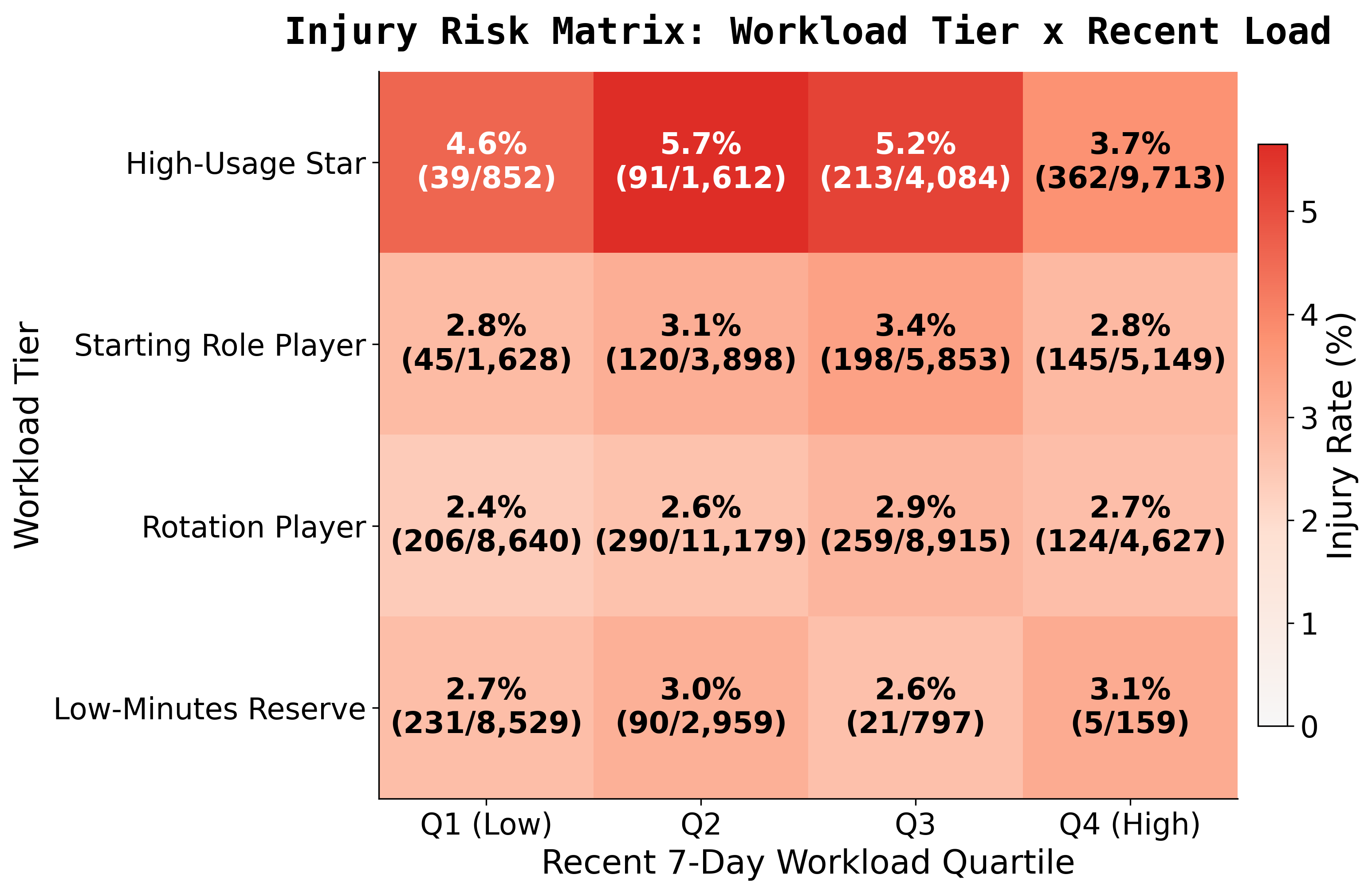}
  \caption{Injury rate (\%) by workload tier and recent 7-day load quartile. High-Usage Stars exhibit a paradoxical inverted gradient (higher load corresponds to lower injury rate), the direct tier-level manifestation of selection bias.}
  \label{fig:risk_matrix}
\end{figure}

However, the paradox is most striking among High-Usage Stars. As illustrated in Figure~\ref{fig:risk_matrix}, this cohort exhibits an \textbf{inverted load gradient}: per-game injury rates actually \emph{decrease} from $4.58\%$ in the lowest recent-load quartile to $3.73\%$ in the highest. Among NBA stars, those with the heaviest recent workloads are precisely those who are healthy enough to sustain them, artificially driving down their observed injury rate.

\section{Discussion and Concluding Remarks}\label{sec:discussion}

\subsection{Resolving the load management paradox}

Naive survival models applied to NBA game-log data consistently produce a dangerous paradox: players who recently logged heavy minutes appear less likely to sustain an injury. In this paper, we have demonstrated that this paradox is not evidence of a physiological protective effect, but rather the mathematical artifact of the healthy-worker survivor effect (HWSE). By formalizing the problem via a longitudinal causal DAG, we identified severe collider bias at the game-participation node: only the most robust, healthy players are selected by coaches to accumulate high workloads.

Our proposed Marginal Structural Piecewise Exponential Model (MS-PEM) provides the first bias-aware survival analysis of NBA injury risk that formally addresses this selection mechanism. By unifying inverse probability of treatment weighting (IPTW) with flexible piecewise-exponential additive models and weighted cumulative exposure (WCE), we offer a methodological template for detecting, quantifying, and partially correcting this bias. 

The IPW correction attenuates the paradoxical WCE weights by $1$--$2\%$ under conservative cross-validated regularization, and by $63$--$78\%$ under lighter penalization across four distinct propensity specifications. This range, supported by an AIPW doubly robust check consistent with adequate outcome-model specification, indicates that while the exact magnitude of the correction remains sensitive to regularization, the \emph{direction} of the correction is highly stable. As confirmed by our simulation study in Section~\ref{sec:simulation}, the HWSE is entirely sufficient to reverse the sign of a true positive workload--injury association, and adjusting for observed selection reliably shifts the hazard back toward the true underlying risk.

\subsection{The structural nature of the bias across sports}

The implications of this paradox extend far beyond professional basketball. \emph{Any} observational analysis of workload and injury that conditions on athlete participation is highly susceptible to this structural bias. This encompasses workload studies in soccer \citep{zumetaolaskoaga2025pamm}, rugby \citep{williams2022cumulative}, and cricket, as well as the vast literature relying on the Acute:Chronic Workload Ratio (ACWR) \citep{gabbett2016paradox} computed strictly from played-game data. 

Because this bias arises directly from the causal graph topology rather than from sample size limitations or functional form misspecification, standard predictive models---including highly flexible machine learning approaches---will reliably learn and amplify this artifact if they are not explicitly constrained by a causal framework.

\subsection{Limitations of observational correction}

The most consequential limitation of this framework is that IPW only corrects for selection on \emph{observed} covariates; it cannot entirely eliminate bias emanating from unobserved latent fitness. Because the exact fraction of selection captured by observed covariates cannot be perfectly determined from game logs alone, we view the MS-PEM as providing a \emph{partially corrected} estimate. Our simulation study highlights this limit: when latent fitness also dictates continuous exposure variations (e.g., minute restrictions within a game), IPW provides incomplete correction even when the binary selection confounder is known.

Furthermore, our injury definition inherits the coarseness of official NBA injury reports \citep{mack2019database}. Minor complaints that do not result in an ``Out'' status go unrecorded, and prophylactic rest absences may occasionally mask underlying micro-traumas. We also treat all injuries as a single outcome class, despite acute contact injuries and chronic overuse conditions likely following different etiological pathways \citep{torresronda2022epidemiology}. Finally, the workload tier classification is constructed from season-level summaries, making it a post-baseline, exposure-contaminated construct rather than a clean baseline covariate; this must be kept in mind when interpreting tier-stratified results.

\subsection{Practical implications and future directions}

For NBA front offices and sports science staffs, the central takeaway is highly actionable: workload--injury models that blindly condition on game participation will systematically underestimate injury risk for high-workload players. This is most acutely visible among High-Usage Stars, who exhibit an inverted load gradient---the clearest tier-level evidence of the paradox. Relying on naive models may lead teams to falsely conclude that heavy workloads are ``safe'' for their most valuable players, when in reality, those workloads are simply a marker of peak baseline fitness.

Fully deconfounding the workload--injury relationship to optimize rest-day policies requires moving beyond observational game logs. Bridging the remaining gap to full causal identification will require:
\begin{enumerate}[nosep]
  \item \textbf{Wearable sensor data}: Continuous physiological monitoring (e.g., heart rate variability, accelerometry) that captures latent fitness directly, closing the unobserved confounding gap.
  \item \textbf{Instrumental variables}: Sources of exogenous variation in workload (e.g., schedule randomizations, acute travel disruptions) unrelated to injury risk through other paths.
  \item \textbf{Negative controls}: Utilizing exposure and outcome variables known to be causally null to benchmark residual confounding \citep{hernan2020whatif}.
\end{enumerate}

Ultimately, sports analytics must transition from predictive associations to causal realities. Methodologies like the MS-PEM provide the necessary first step, demonstrating that without acknowledging playing time as a consequence of health, we cannot accurately measure it as a cause of injury.

\section*{Code availability.}
All code, data collection scripts, and analysis notebooks are available at \url{https://github.com/khrisyu9/nba-load-management}.


\bibliographystyle{ims}
\bibliography{reference}

\newpage


\appendix

\section{Additional Figures}\label{app:figures}
We present additional figures in Appendix~\ref{app:figures} to illustrate details about data and our methods.
\begin{figure}[H]
  \centering
  \includegraphics[width=0.85\textwidth]{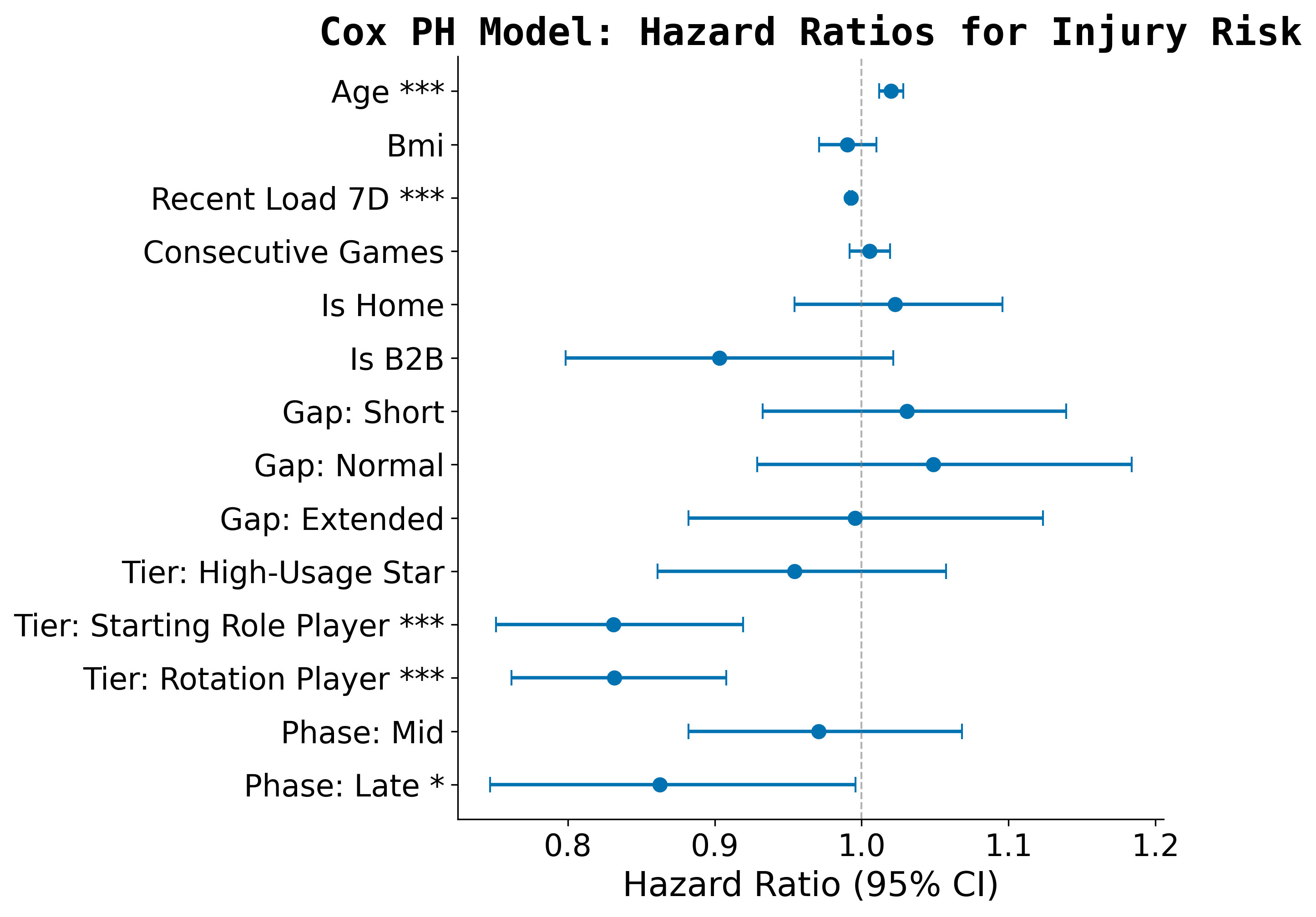}
  \caption{Forest plot of Cox proportional hazards ratios with 95\% confidence intervals.}
  \label{fig:forest}
\end{figure}

Figure~\ref{fig:forest} visually summarizes the baseline Cox results and makes the paradox especially clear. Age is the most consistent positive predictor of injury risk, while recent 7-day load enters with a statistically precise hazard ratio below one. We do not interpret that negative recent-load coefficient as evidence that heavier workloads are protective; rather, it is the same paradoxical pattern motivating the paper, consistent with healthy-worker survivor selection. The figure also shows that Starting Role Players and Rotation Players have lower estimated hazard than Low-Minutes Reserves, whereas most game-gap indicators remain comparatively close to the null with wider uncertainty.

\begin{figure}[H]
  \centering
  \includegraphics[width=0.85\textwidth]{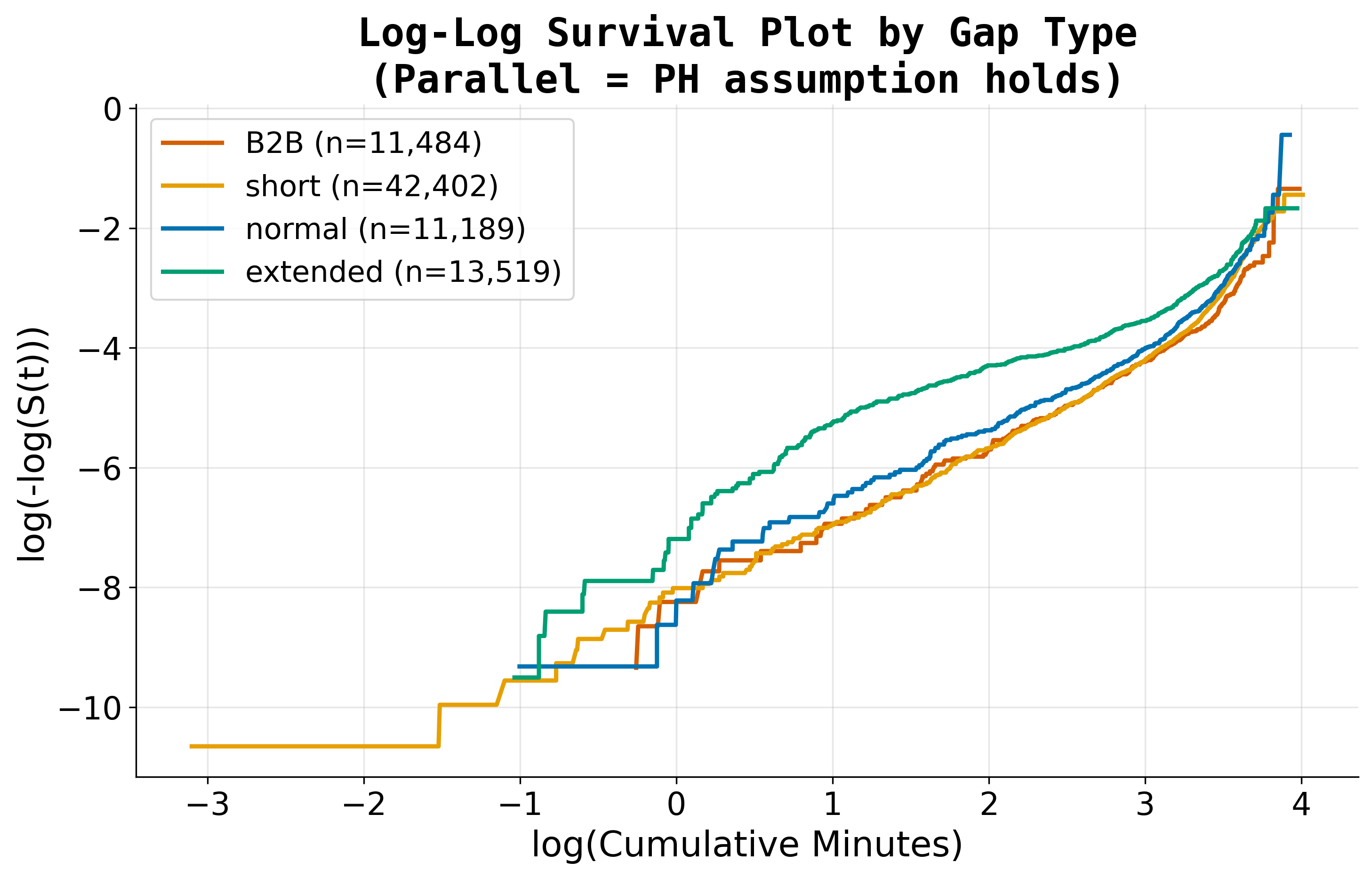}
  \caption{Log-log survival plots by game-gap type, showing violation of the proportional hazards assumption.}
  \label{fig:loglog}
\end{figure}

Figure~\ref{fig:loglog} shows that if the proportional hazards assumption held exactly, the log-log curves would be approximately parallel. Instead, the four gap-type curves clearly diverge and re-converge over cumulative minutes, especially for the extended-rest group in the middle portion of the curve. This indicates that the effect of schedule spacing is not constant over the exposure history, which helps justify moving from a standard Cox model to the more flexible piecewise-exponential framework used later in the paper.

\begin{figure}[H]
  \centering
  \includegraphics[width=0.75\textwidth]{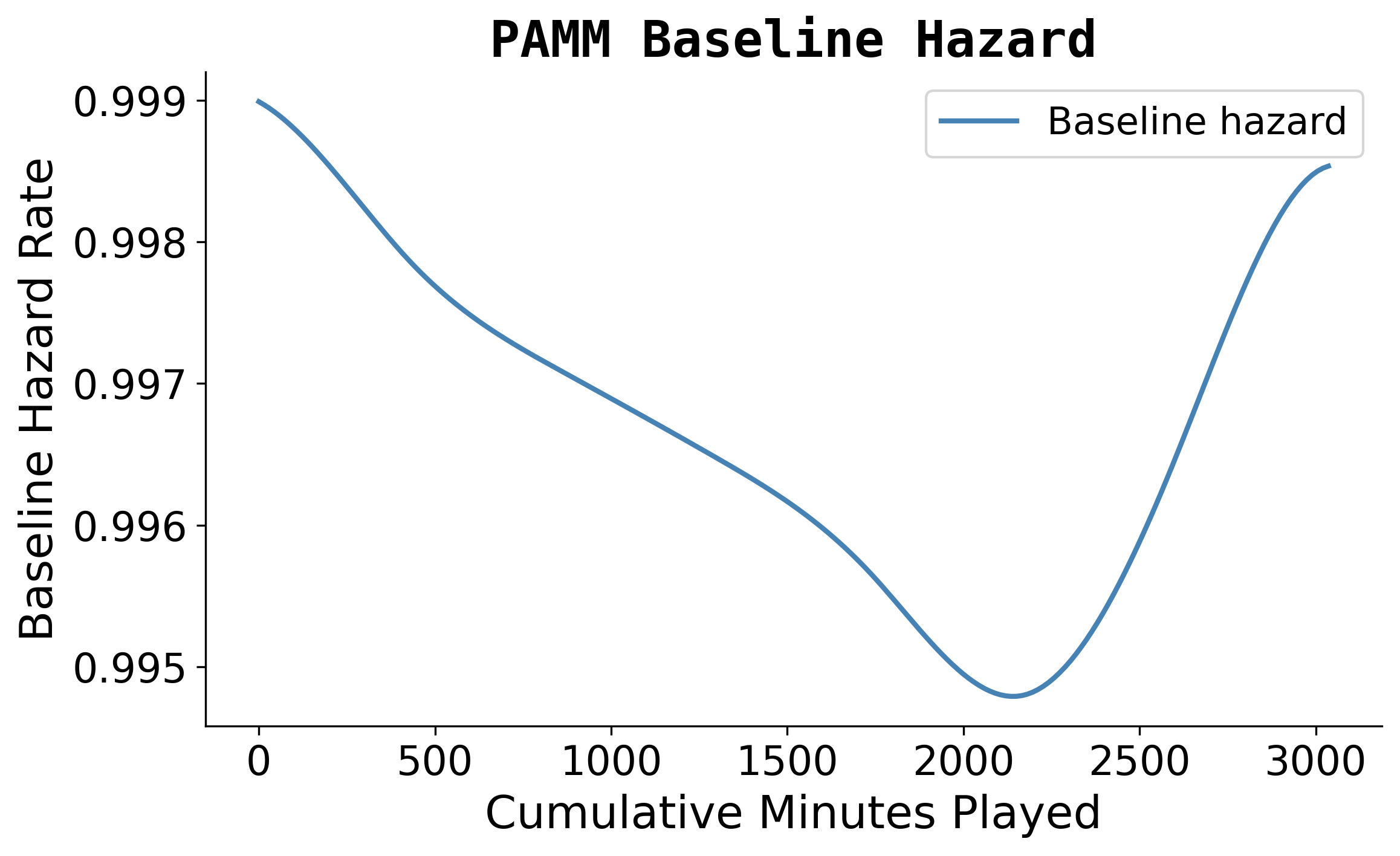}
  \caption{Smooth baseline hazard from the piecewise-exponential model. The non-monotonic shape reflects both fatigue accumulation and survivorship selection.}
  \label{fig:baseline}
\end{figure}

Figure~\ref{fig:baseline} show that the estimated baseline hazard is distinctly non-monotone: it declines through much of the cumulative-minutes range and then rises again late in the exposure history. A natural interpretation is that these two phases combine survivor selection and physiological accumulation. Early in the season, the risk set becomes increasingly enriched for players healthy enough to keep playing, which can push the baseline hazard downward; later, once cumulative exposure becomes large, accumulated workload appears to push the hazard back up. This shape is exactly the kind of pattern that a rigid parametric or proportional-hazards baseline would struggle to represent.

\begin{figure}[H]
  \centering
  \includegraphics[width=0.95\textwidth]{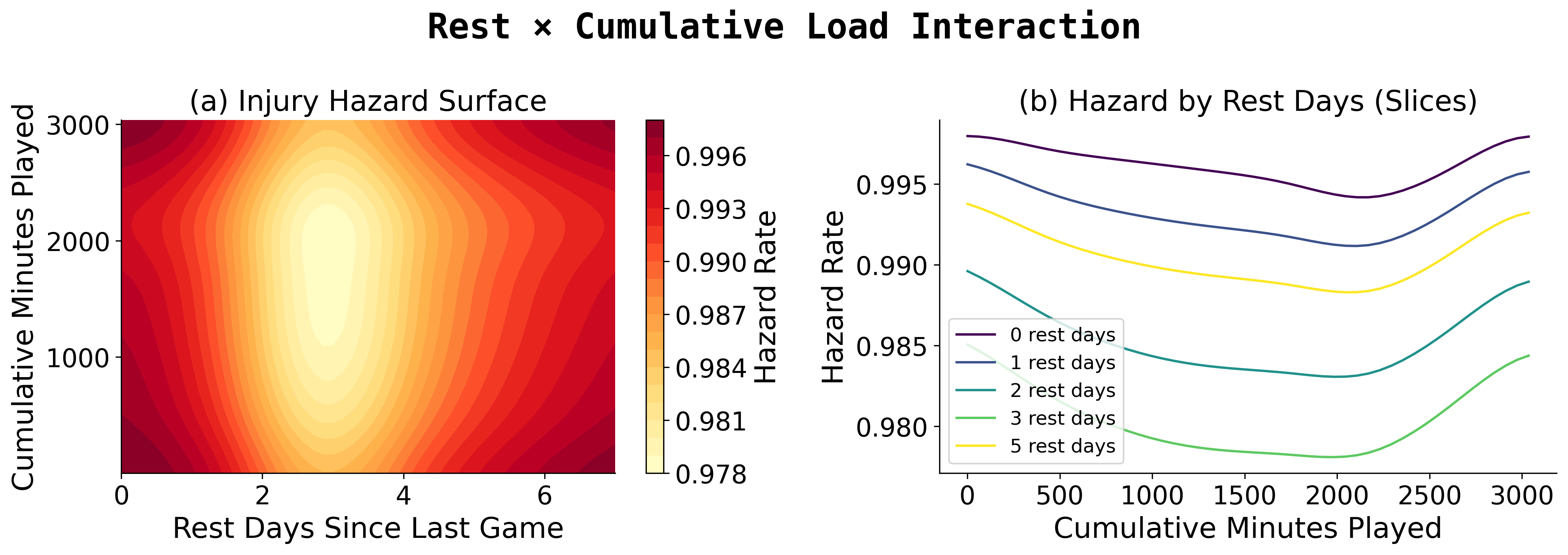}
  \caption{Hazard surface from the tensor-product interaction of rest days and cumulative minutes in the PAMM baseline model.}
  \label{fig:surface}
\end{figure}

The hazard surface plot, Figure~\ref{fig:surface} (a), shows that injury risk varies jointly across rest days and cumulative minutes rather than along either axis alone. The slice plot, Figure~\ref{fig:surface} (b), suggests that shorter rest tends to shift the hazard upward across most cumulative-minute levels, while all curves retain a non-monotone shape over cumulative minutes, with risk dipping in the middle range and rebounding at very high totals. Importantly, it should be read as a descriptive summary from the baseline PAMM: it captures complex structure in the observed data, but it does not by itself separate genuine fatigue effects from the selection mechanism emphasized throughout the paper.

\begin{figure}[H]
  \centering
  \includegraphics[width=0.85\textwidth]{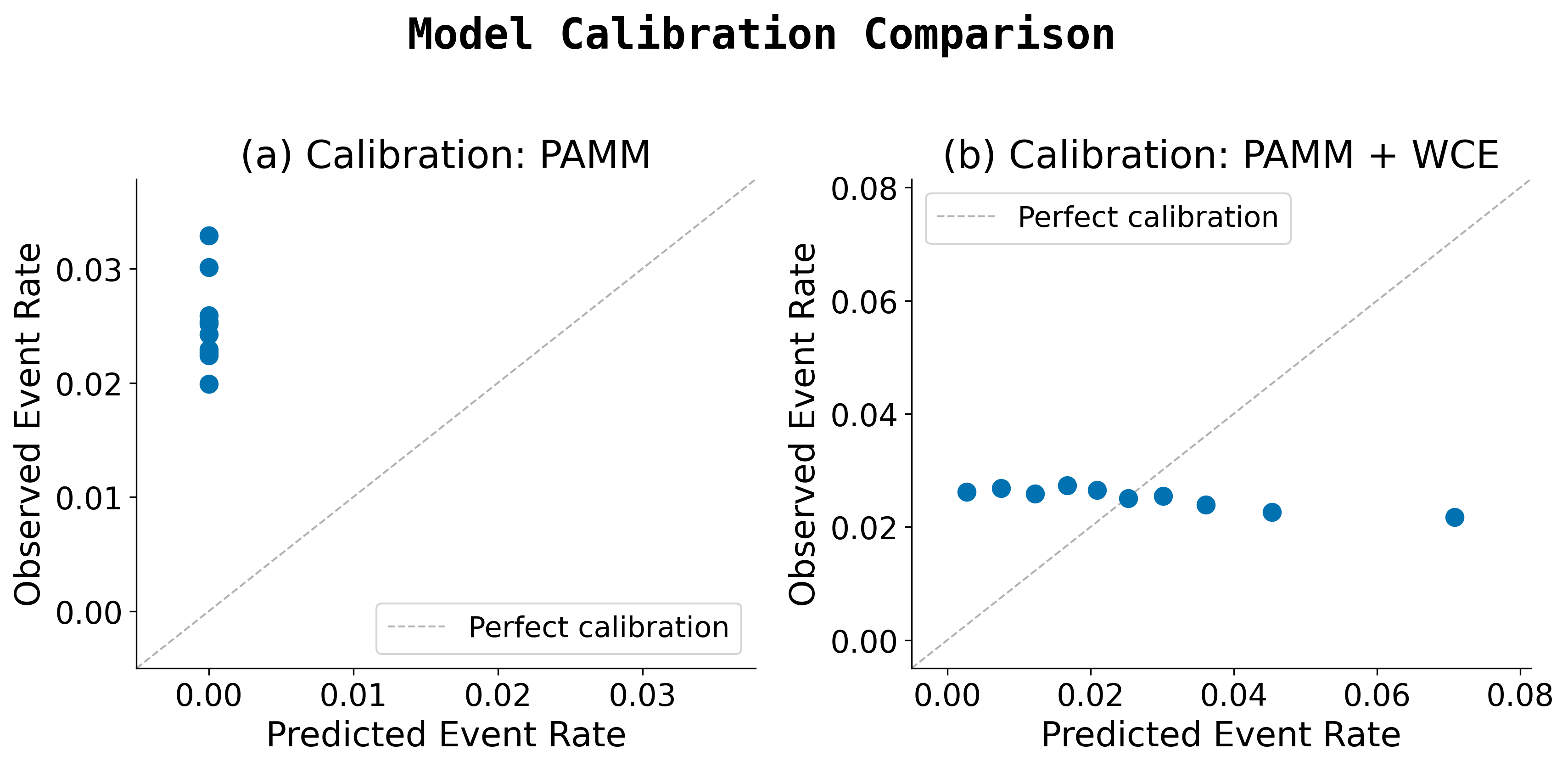}
  \caption{Calibration plot comparing predicted and observed event rates by decile.}
  \label{fig:calibration}
\end{figure}

Figure~\ref{fig:calibration} is a useful calibration diagnostic because it highlights both the improvement achieved by the model and its remaining limitations. The plain PAMM tends to underpredict event risk almost uniformly, with predicted probabilities clustered near zero while observed event rates remain around two to three percent across deciles. Adding the WCE term produces more spread in predicted risk and better separates low- and high-risk bins, but the points still do not fall tightly on the 45-degree line, especially in the highest predicted-risk deciles. Accordingly, these models are more informative about relative hazard structure and bias patterns than about perfectly calibrated absolute injury probabilities.

\begin{figure}[H]
  \centering
  \includegraphics[width=0.85\textwidth]{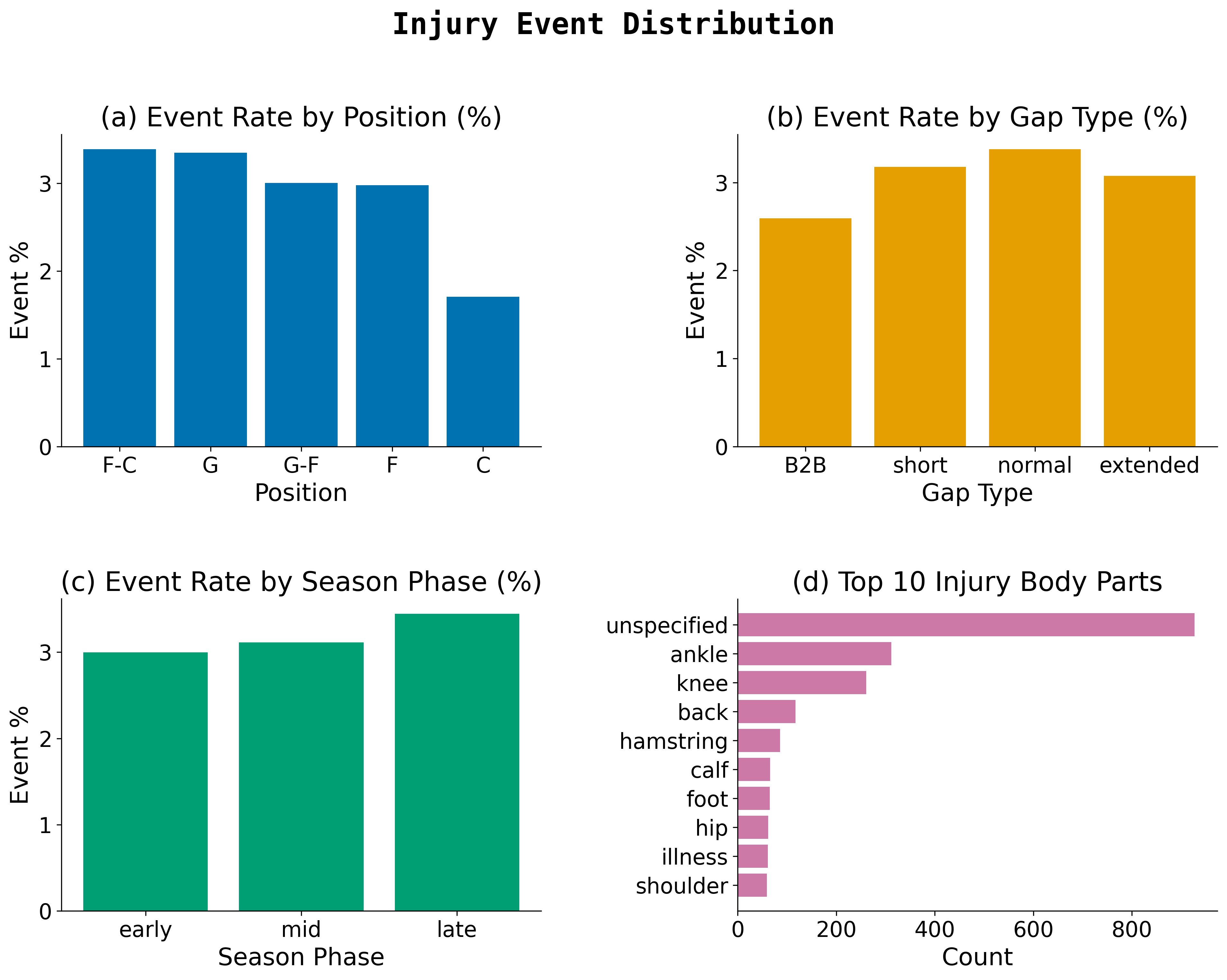}
  \caption{Distribution of injury events by position, season phase, and gap type.}
  \label{fig:event_dist}
\end{figure}

Several descriptive patterns in Figure~\ref{fig:event_dist} match the paper's broader argument. First, back-to-back games again show the lowest raw event rate, while short and normal gaps appear riskier, which is another unadjusted manifestation of the load-management paradox rather than evidence that compressed schedules are protective. Second, event rates rise modestly toward the late season, consistent with cumulative wear over time. Third, the body-part panel shows that ankle and knee injuries are among the most common named categories, but the large ``unspecified'' group also highlights a limitation of public injury reports: the observed outcome pool is etiologically heterogeneous, so the paper's single-outcome hazard model should be interpreted as a useful aggregate approximation rather than a mechanism-specific injury model.

\begin{figure}[H]
  \centering
  \includegraphics[width=0.85\textwidth]{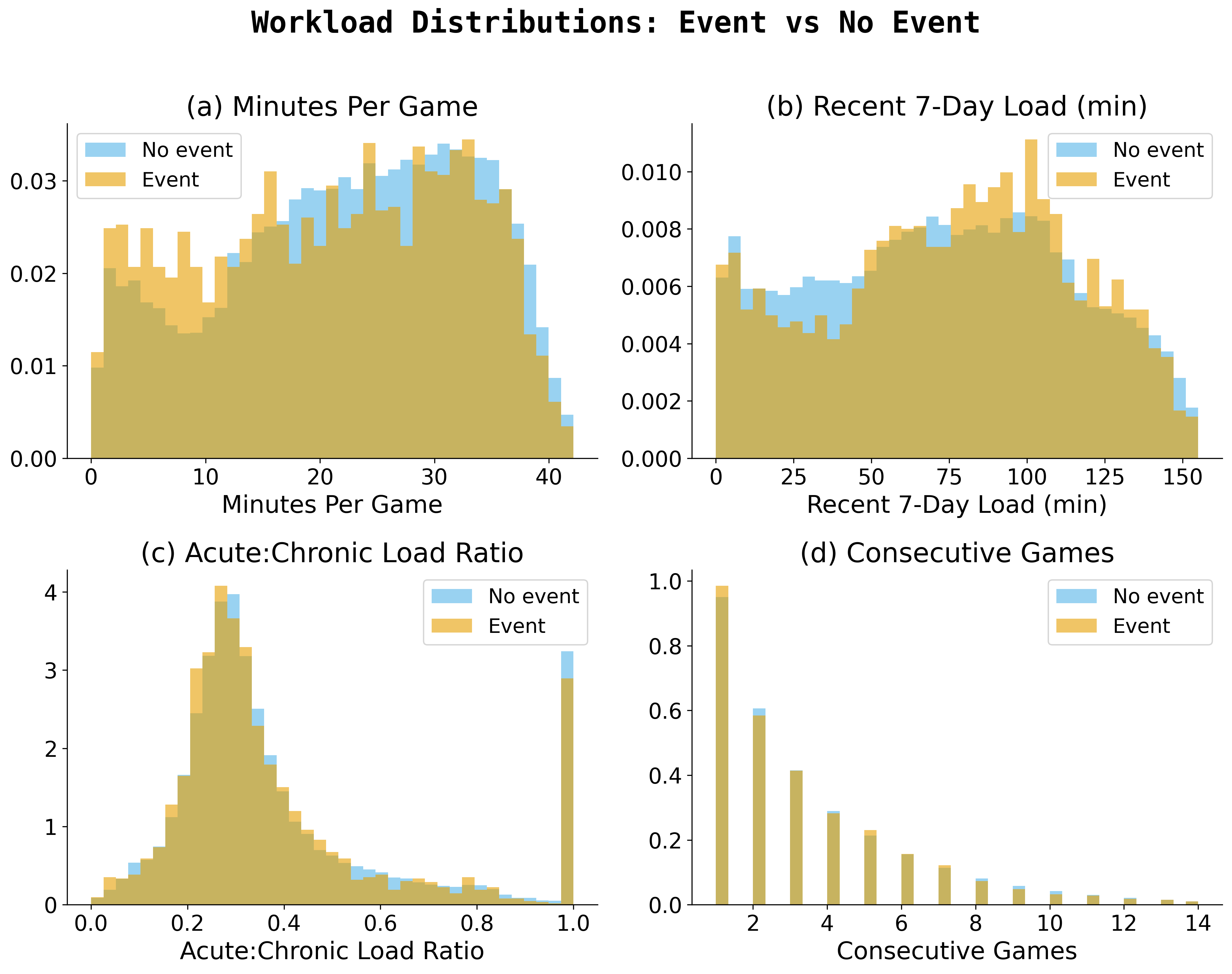}
  \caption{Distribution of key workload variables across the combined dataset.}
  \label{fig:workload}
\end{figure}

From Figure~\ref{fig:workload}, we can see that the event and no-event distributions overlap heavily for all four workload summaries, which is an important practical point: no single threshold in minutes, recent load, acute:chronic ratio, or consecutive games cleanly separates injury from non-injury observations. This is one reason simple cutoff-based rules can be misleading in this setting. The figure instead supports the modeling strategy of the paper, where risk is treated as longitudinal, cumulative, and multi-variable rather than reducible to a one-dimensional screening rule.

\begin{figure}[H]
  \centering
  \includegraphics[width=0.75\textwidth]{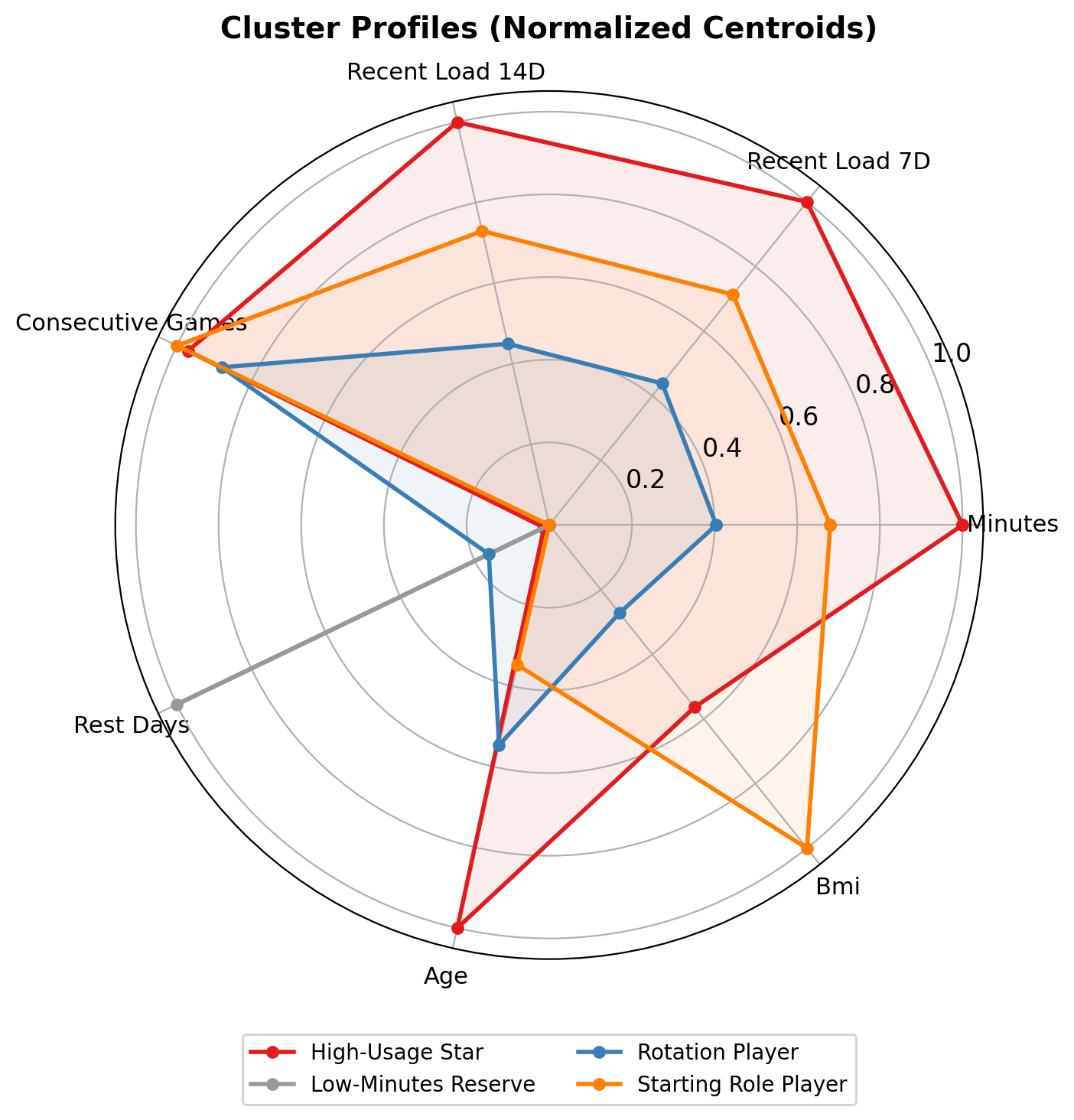}
  \caption{Player workload cluster profiles.}
  \label{fig:cluster}
\end{figure}

The radar plot, Figure~\ref{fig:cluster}, confirms that the clustering procedure recovers interpretable role-based workload profiles. High-Usage Stars sit at the top of the minutes and recent-load axes, Starting Role Players remain high but below stars, Rotation Players occupy the middle range, and Low-Minutes Reserves are characterized by little sustained workload and the most rest days. This is useful for descriptive stratification because it shows that ``workload tier'' is not merely relabeled minutes played; it also reflects continuity of play and exposure accumulation. At the same time, these tiers are constructed from season-level summaries, so they should be read as descriptive player archetypes rather than clean baseline causal covariates.

\begin{figure}[H]
  \centering
  \includegraphics[width=0.95\textwidth]{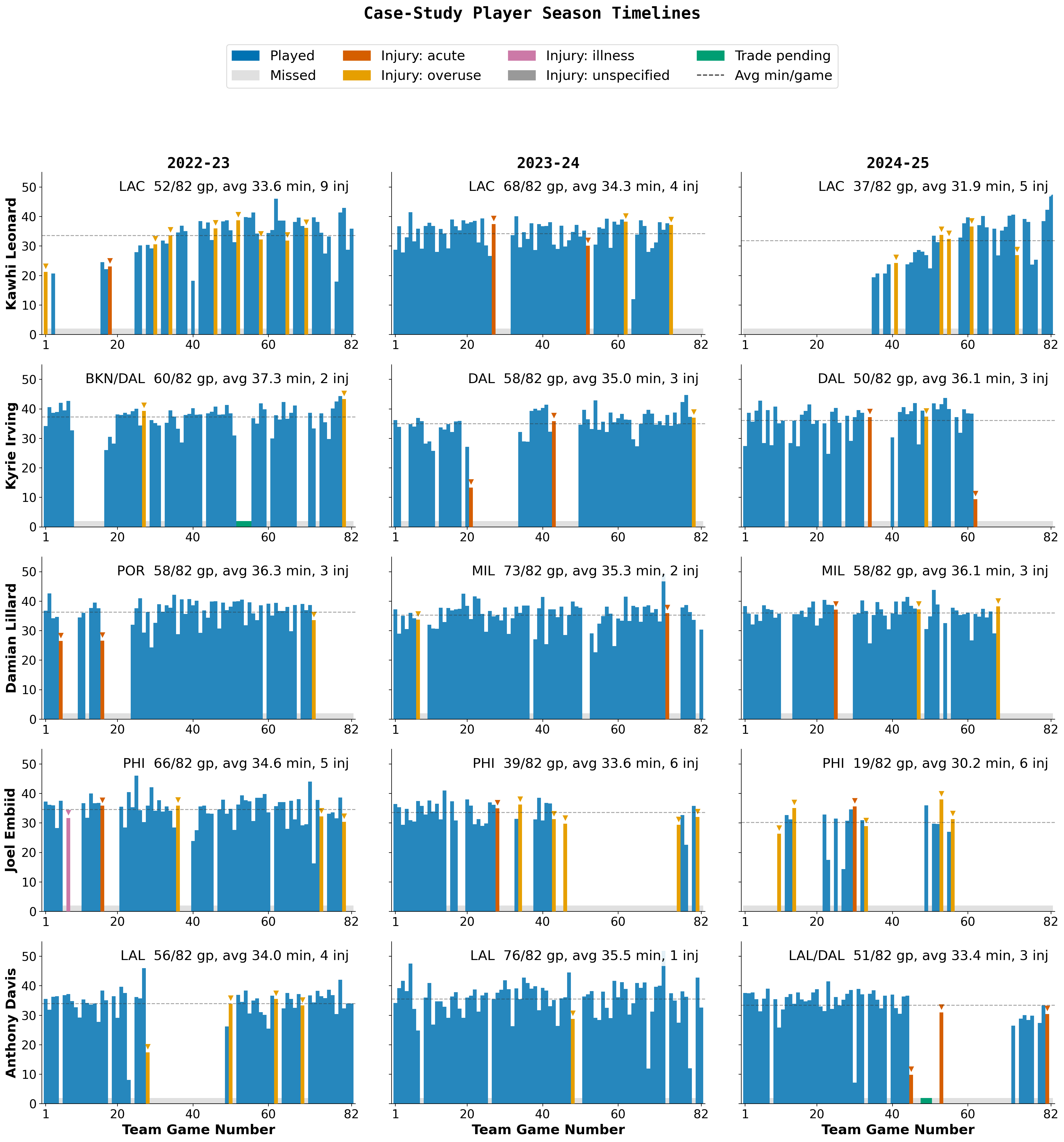}
  \caption{Season timelines for five case-study players across three seasons.}
  \label{fig:timelines}
\end{figure}

In Figure~\ref{fig:timelines}, we present the injury histories of five case-study star players who suffered with high-frequent or devastating injuries over the 2022--2025 seasons. These timelines make the selection problem visually intuitive. In general, extended stretches of high-minute games are observed only when a player remains healthy and available enough to stay in the rotation, whereas injuries and missed periods interrupt the exposure history in highly player-specific ways. The figure also reveals substantial heterogeneity across both players and seasons, further motivating the paper’s use of recurrent-event methods and time-varying covariates. For example, Kawhi Leonard, who is widely associated with load management, shows repeated appearances of ``injury: overuse,'' often followed by short absences over the three seasons, underscoring the practical importance of workload monitoring in his case. Kyrie Irving provides a different pattern: despite experiencing a lumbar back sprain during the 2024--25 season, he continued to play under a dense game schedule until sustaining a season-ending left ACL tear later in the year. Anthony Davis, by contrast, missed a substantial number of games around the midseason trade to Dallas and then suffered a left adductor strain in his Mavericks debut, which sidelined him for more than a month afterward. More broadly, these case studies of several star players illustrate that observed workload is not purely an exogenous driver of future injury risk; it is also shaped by a player’s evolving health status and availability.

\begin{figure}[H]
  \centering
  \includegraphics[width=0.95\textwidth]{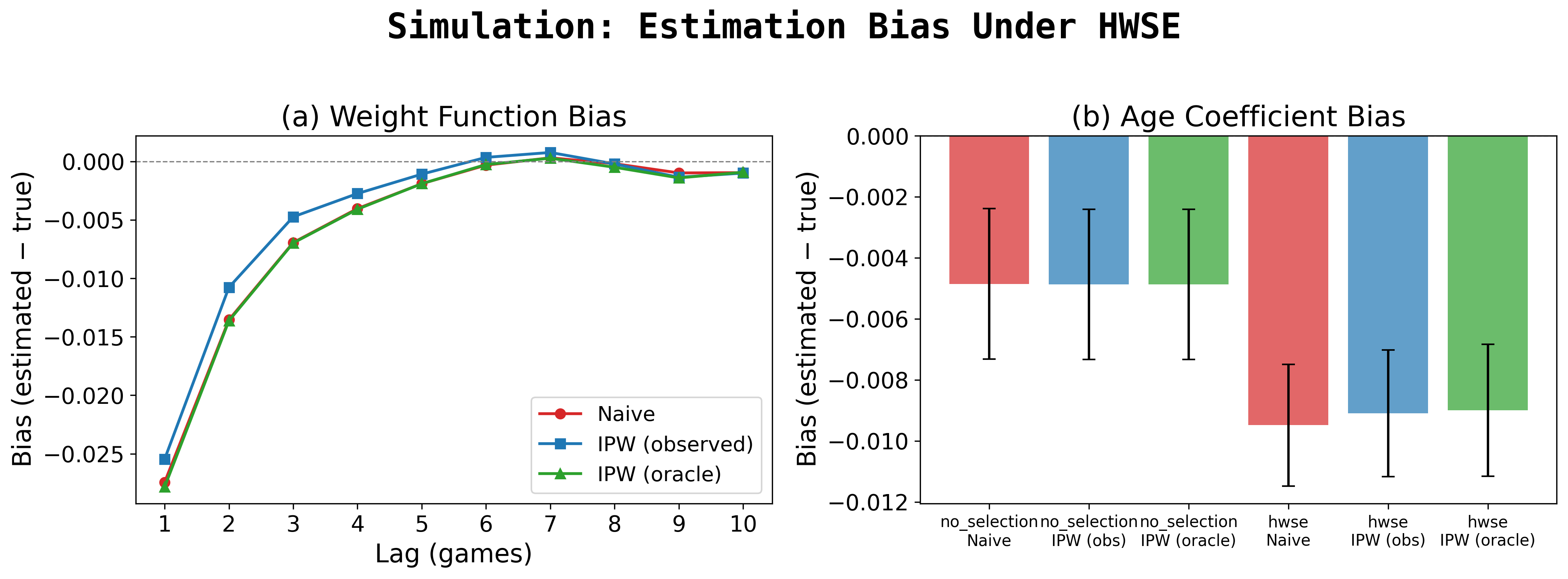}
  \caption{Simulation: estimation bias under HWSE. (a) Weight function bias at each lag for naive, IPW-observed, and IPW-oracle estimators. (b) Age coefficient bias across scenarios and models.}
  \label{fig:sim_bias}
\end{figure}

Figure \ref{fig:sim_bias} clarifies where the healthy-worker survivor effect does the most damage. In panel (a), the bias is most severe at the shortest lags, exactly where recent workload should matter most biologically, and then shrinks toward zero at longer lags. The observed-covariate IPW estimator attenuates the short-lag bias but does not eliminate it, showing that weighting on publicly available variables provides only a partial correction. Panel (b) further shows that the distortion is not limited to the WCE curve: the same selection mechanism can also bias ordinary regression coefficients such as age. Taken together, the simulation supports the paper's main empirical message that the causal correction moves the estimates in the right direction, but cannot fully recover the truth when latent fitness remains unobserved.

\section{E-Value Sensitivity Analysis}\label{app:evalue}

Following \citet{vanderweele2017evalue}, we compute E-values for key Cox hazard ratios, and the results are shown in Table~\ref{tab:evalue}.
For $\text{HR} < 1$, $E\text{-value} = \text{HR}^{-1} + \sqrt{\text{HR}^{-1}(\text{HR}^{-1} - 1)}$.
The continuous per-minute recent-load E-value ($1.09$) is scale-dependent: very weak confounding per minute suffices to explain the association.

\begin{table}[H]
\centering
\caption{E-value sensitivity analysis for key hazard ratios.}
\label{tab:evalue}
\begin{tabular}{lccc}
\toprule
Covariate & HR & E-value (point) & E-value (CI bound) \\
\midrule
Recent load (7d)       & 0.993 & 1.09 & 1.08 \\
Tier: Starting Role    & 0.831 & 1.70 & 1.40 \\
Tier: Rotation Player  & 0.832 & 1.69 & 1.44 \\
Gap: short             & 1.031 & 1.21 & 1.35 \\
\bottomrule
\end{tabular}
\end{table}

\section{Player Workload Tier Clustering}\label{app:clustering}

Players are classified into workload tiers using the following pipeline, applied separately to each season's game logs:

\begin{enumerate}
  \item \textbf{Feature aggregation.} For each player with at least $10$ games, we compute $14$ per-season summary statistics spanning five domains: \emph{volume} (mean minutes, field-goal attempts, rebounds, personal fouls per game), \emph{availability} (games played, fraction of games with $\geq 30$ minutes), \emph{usage} (usage rate), \emph{playmaking} (mean assists), \emph{defense/physicality} (mean steals, blocks, offensive rebounds, free-throw attempts), and \emph{efficiency} (true shooting percentage).
  \item \textbf{Standardization and PCA.} Features are $z$-scored, then reduced via principal component analysis \citep{jolliffe2016pca} retaining $85\%$ of total variance.
  \item \textbf{$K$-means clustering.} We fit $K$-means \citep{lloyd1982kmeans} on the PCA scores for $k \in \{3, \ldots, 7\}$ and select the $k$ maximizing the silhouette score \citep{rousseeuw1987silhouettes}. All three seasons yield $k = 4$.
  \item \textbf{Labeling.} Clusters are ranked by centroid mean minutes and assigned interpretable names: \emph{High-Usage Star} (highest minutes and usage rate), \emph{Starting Role Player}, \emph{Rotation Player}, and \emph{Low-Minutes Reserve}. Players with fewer than $10$ games are assigned to the Low-Minutes Reserve tier by default.
\end{enumerate}
\noindent Figure~\ref{fig:cluster}, the radar plot, shows the normalized centroid profiles for the four tiers.

\end{document}